\documentclass[fleqn,usenatbib]{mnras}

\usepackage{newtxtext,newtxmath}
\usepackage{gensymb}

\usepackage[T1]{fontenc}

\DeclareRobustCommand{\VAN}[3]{#2}
\let\VANthebibliography\thebibliography
\def\thebibliography{\DeclareRobustCommand{\VAN}[3]{##3}\VANthebibliography}

\usepackage{graphicx}	% Including figure files
\usepackage{amsmath}	% Advanced maths commands
\usepackage{xcolor}
\newcommand{\src}{V1674~Her}

\usepackage{hyperref}
\title[Optical analysis of V1674 Her]{Optical observations of the fast nova V1674 Herculis}

\author[Rawat et al.]{
Neeraj Singh Rawat$^{1,}$$^{2}$\thanks{E-mail: \href{mailto:neeraj.singh@iiap.res.in}{neeraj.singh@iiap.res.in}; \href{mailto:neerajrawat377@gmail.com}{neerajrawat377@gmail.com}},
L. S. Sonith$^{1,}$$^{2}$,
U. S. Kamath$^{1,}$$^{2}$,
Yash Bhargava$^{3}$,
G. C. Anupama$^{1}$, and\newauthor
Kulinder Pal Singh$^{4}$
\\
$^{1}$Indian Institute of Astrophysics, 2nd Block, Koramangala, Bengaluru 560034, India\\
$^{2}$Pondicherry University, R.V. Nagar, Kalapet, Puducherry 605014, India\\
$^{3}$INAF–Osservatorio Astronomico di Cagliari, Via della Scienza 5, 09047 Selargius (CA), Italy\\
$^{4}$Department of Physical Sciences, Indian Institute of Science Education and Research Mohali, Knowledge City, Sector 81, SAS Nagar, Punjab 140306, India
}

\date{Accepted XXX. Received YYY; in original form ZZZ}

\pubyear{2026}

\begin{document}
\label{firstpage}
\pagerange{\pageref{firstpage}--\pageref{lastpage}}
\maketitle

\begin{abstract}
We present the evolution of optical spectra and lightcurves of the fast nova V1674~Herculis during 150 days past its eruption. Using the post-eruption AAVSO light curve, we have calculated the orbital period of V1674~Her to be 0.153 days. There is no unambiguous white dwarf spin period in our data. The optical spectra show that the ionisation increases with time. A morpho-kinematic analysis of the H$\alpha$ line profile indicates a bipolar morphology with polar blobs and an equatorial ring. Lyman beta fluorescence is found to be the dominant mechanism for the excitation of neutral oxygen. On day 19.87, [Ne III] \& [Ne V] lines are present, indicating the presence of the ONe white dwarf. On day 147.66, the nebular lines are still present, implying that the nova had not gone into quiescence yet; this spectrum is accretion-dominated. 
\end{abstract}

\begin{keywords}
transients: novae -- techniques: photometric -- techniques: spectroscopic -- stars: individual: V1674 Herculis
\end{keywords}

%%%%%%%%%%%%%%%%% BODY OF PAPER %%%%%%%%%%%%%%%%%%

\section{Introduction}
\subsection{Classical Novae}
Classical novae (CNe) are explosive events that occur in binary star systems consisting of a white dwarf (WD) and a companion star that fills its Roche Lobe. These systems are known as cataclysmic variables and are characterised by the transfer of matter from the companion star through the inner Lagrangian point onto the surface of the WD via an accretion disk \citep{1995cvs..book.....W}. In these close binary systems, the WD and the companion star orbit each other with a period of a few hours. When enough material accumulates on the WD, it can trigger a thermonuclear runaway (TNR) in the matter that is electron degenerate \citep{2012BASI...40..419S, Starrfield2020ApJ...895...70S}. As a result, an envelope of the accreted material gets ejected, with ejecta masses that can range from $\sim10^{-7}$ to $10^{-4}$ \(\textup{M}_\odot\), depending primarily on the WD mass, its core temperature, and the mass accretion rate \citep{2005ApJ...623..398Y}, and with velocities ranging from a few 100 km/s to several 1000 km/s, leading to a sudden increase in the optical brightness of the system by  $\sim11$ magnitudes \citep{2021ApJ...910..120K} - this is known as a nova eruption. Following a nova eruption, the spectrum exhibits signatures of rapidly expanding ejecta and ongoing thermonuclear processes on the WD surface \citep{1992AJ....104..725W}. In the earliest phase of a nova eruption, the spectra are dominated by permitted recombination lines due to low ionisation. As the ejecta expand, deeper and hotter regions near the white dwarf become exposed, resulting in a rise in ionisation and the emergence of high-ionisation forbidden and coronal features. The entire eruption phase can span from weeks to several months, after which the system gradually fades back to its pre-eruption state. As the nova settles into quiescence, the ionisation level decreases once again (for more details, see \citet{2012BASI...40..161A} and references therein).

\subsection{V1674 Her: Discovery and Peculiar Characteristics}
V1674 Her (Nova Her 2021; TCP J18573095+1653396) was discovered in eruption by Seĳi Ueda (Kushiro, Hokkaido, Japan)\footnote{\url{http://www.cbat.eps.harvard.edu/unconf/followups/J18573095+1653396.html}} at 8.4 magnitudes on 2021 June 12.537 UT. V1674~Her is the fastest galactic nova ever with a $t_2$ (time taken to decline by two magnitudes from the maximum) value of 1.2 days \citep{2021ApJ...922L..10W}. Adding to its intrigue, \citet{2021RNAAS...5..160Q} noted a pre-eruption plateau in V1674~Her's light curve at around g $\sim14$ magnitudes and also pointed out the fact that while such occurrence of a plateau has been reported for a few cases, the detection at $\sim8$ magnitudes below peak is unprecedented in their knowledge. Utilising the pre-eruption archival ZTF light curve of the nova, \citet{2021ATel14720....1M} found a period of 501.4277 s and interpreted it as the spin period of the WD in an intermediate polar system. Subsequent observations post-eruption with NICER \citep{2021ATel14798....1P} and Chandra \citep{2021ATel14776....1M} reported a period of 501.8 ± 0.7 s and 503.9 s, respectively. Notably, \citet{2022ApJ...940L..56P} found a spin period of 501.486(5) s and additionally identified a signal at 0.152921(3)d, indicating it to be the orbital period of the WD. \citet{2021ATel14835....1S} reported a similar orbital period based on optical data. \citet{2024BAAA...65...60L}, using TESS observations, detected a 0.153d period consistent with the orbital period of the system. They also reported an additional longer periodicity near $\sim0.537$d, of unknown origin. \citet{2022MNRAS.517L..97L} also found a similar orbital period of 0.153d based on NICER observations. \citet{2021ApJ...922L..42D} found a spin period of 501.72 ± 0.11 s based on Chandra observations, while \citet{2024MNRAS.528...28B} reported a period of 501.4 - 501.5 s in X-rays at the peak of the super-soft source (SSS) phase with AstroSat observations.

Following its eruption, V1674~Her garnered substantial attention and was extensively observed across multiple wavelengths. Initially displaying Fe II lines \citep{2021ATel14723....1W}, and references therein), V1674~Her was classified as a Fe II-type nova \citep{1992AJ....104..725W, 2012AJ....144...98W}. Notably, \citet{2021ATel14728....1W} observed a shift in the near-infrared (NIR) spectra towards the He/N class within 5.5 days after the eruption, suggesting the possibility of V1674~Her being a hybrid nova. Approximately 17 days after the nova eruption, the appearance of the [Ne V] 342.6 nm and [Ne III] 386.9 nm, 396.8 nm lines were reported \citep{2021ATel14746....1W}, indicating that V1674~Her has ONe WD. Additionally, \citet{2021ATel14741....1W} reported the detection of NIR coronal lines on day 11.5, marking the earliest detection of such lines in a classical nova, even preceding the onset of the SSS phase on day 18.9 \citep{2021ATel14747....1P}. \citet{2024MNRAS.527.1405H} have discussed the optical spectra analysis of V1674~Her over the course of 1 month following the eruption.

We present here the evolution of the optical spectra and light curves of the fast nova V1674~Her over 150 days following its eruption. We describe the observations and data reduction steps in Section \ref{Observations}. The orbital and spin periodicity of V1674~Her is detailed in Section \ref{Period}, followed by a detailed spectral evolution in Section \ref{SpectralEvolution}. Section \ref{Description} provides a description of some prominent emission lines. In Section \ref{Modelling}, we discuss the morpho-kinematic analysis of V1674~Her ejecta using the H$\alpha$  line profile. Finally, we present the results and conclusions in Section \ref{Results}.

\section{Observations}\label{Observations}
\subsection{Photometry}

\subsubsection{GIT}
Photometric observations of \src\ were obtained using the 0.7-m GROWTH-India Telescope\footnote{Global Relay of Observatories Watching Transients Happen (\url{https://www.growth.caltech.edu/}), (\url{https://sites.google.com/view/growthindia/})} \citep[GIT;][]{2022AJ....164...90K} located at the Indian Astronomical Observatory (IAO), Hanle, India. GIT is a fully robotic telescope equipped with 4096 $\times$ 4108 pixels Andor iKon-XL camera and SDSS u$^\prime$, g$^\prime$, r$^\prime$, i$^\prime$ and z$^\prime$ filters. The photometric observations began on 2021 July 21 (MJD $\sim59416.66$) and continued till 2021 October 22 (MJD $\sim59509.60$). \src\ was also observed with high-cadence imaging mode using the SDSS g$^\prime$ filter on 2021 July 21 and the r$^\prime$ filter on 2021 July 22 and 23. Data were reduced and PSF photometry was performed using the GIT pipeline developed by \cite{2022MNRAS.516.4517K}.

\begin{figure*}
\begin{center}
\includegraphics[width=2\columnwidth]{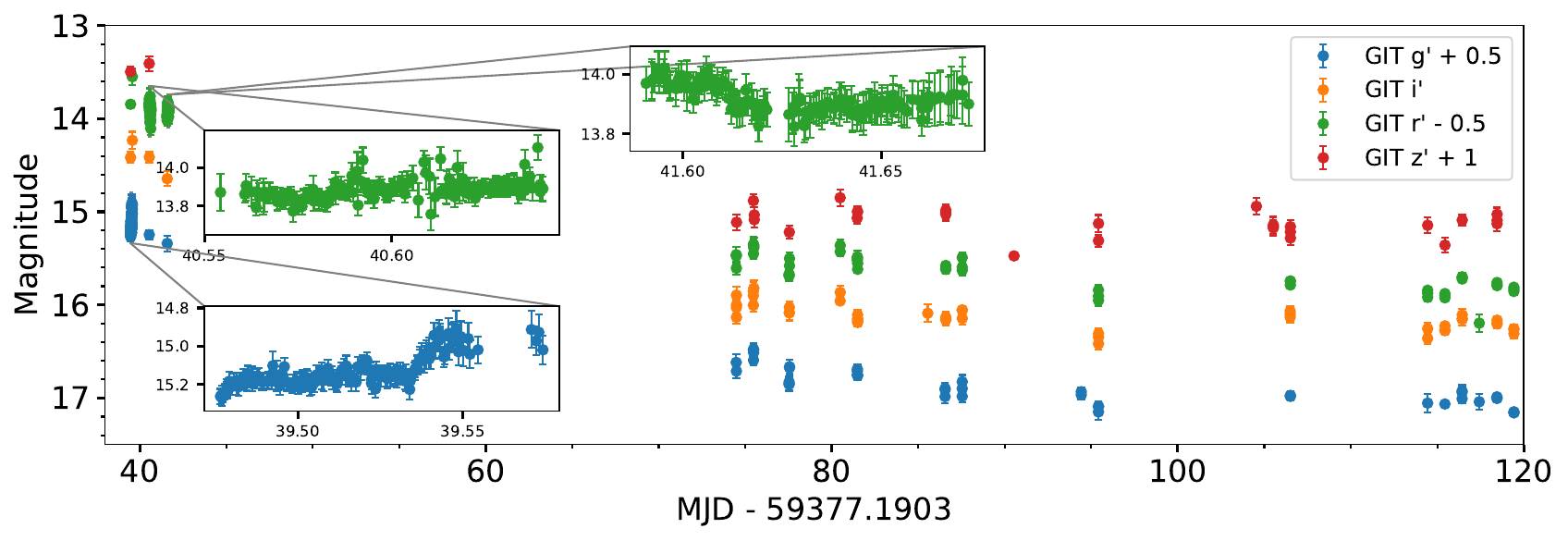}
	\caption{Light curve of nova V1674~Her in the GIT g$^\prime$, r$^\prime$, i$^\prime$ and z$^\prime$ bands. Offset has been applied for individual bands except i$^\prime$ band for better representation. Insets show the high cadence observations.}
	
	\label{git_lc}
\end{center}
\end{figure*}

Details of the observations are summarised in Table \ref{tab:git_log}. Light curve of \src\ constructed using GIT data is shown in Figure \ref{git_lc}.

\begin{table}
\centering
\caption{Log of GIT observations.}
\label{tab:git_log}
\begin{tabular}[h!]{cccc} \hline
Filter & Start time (MJD) & End time (MJD) \\ \hline
g$^\prime$  &  59416.667 & 59509.598 \\
r$^\prime$  &  59416.666 & 59509.599\\ 
i$^\prime$  &  59416.665 & 59509.601\\
z$^\prime$  &  59416.664 & 59509.602\\[1ex]
\multicolumn{1}{c}{High-cadence}\\
\cline{1-1}\\
g$^\prime$  &  59416.667 & 59416.765 \\
r$^\prime$  &  59417.745 & 59417.831\\
r$^\prime$  &  59418.781 & 59418.862 \\ \hline
\end{tabular}
\end{table}

\subsubsection{AAVSO}
The American Association of Variable Star Observers \citep[AAVSO\footnote{\url{https://www.aavso.org}};][]{Kafka2021} International Database compiles observations made primarily by the amateur astronomy community but also includes data from professional researchers. The database has strict quality control measures and is maintained by AAVSO technical staff. We have obtained V- and CV-band data from the AAVSO International Database to find the periodicity of the object. CV corresponds to the unfiltered visual magnitude with Bessel V zeropoint. We have removed data which did not have a ‘V’ (‘observation passed validation tests’) or ‘Z’ (‘pre-validation, data checked for typos and input errors’) flag.

\subsection{Spectroscopy}\label{Spectroscopy}
\begin{figure}
    \centering
    \includegraphics[width=\columnwidth]{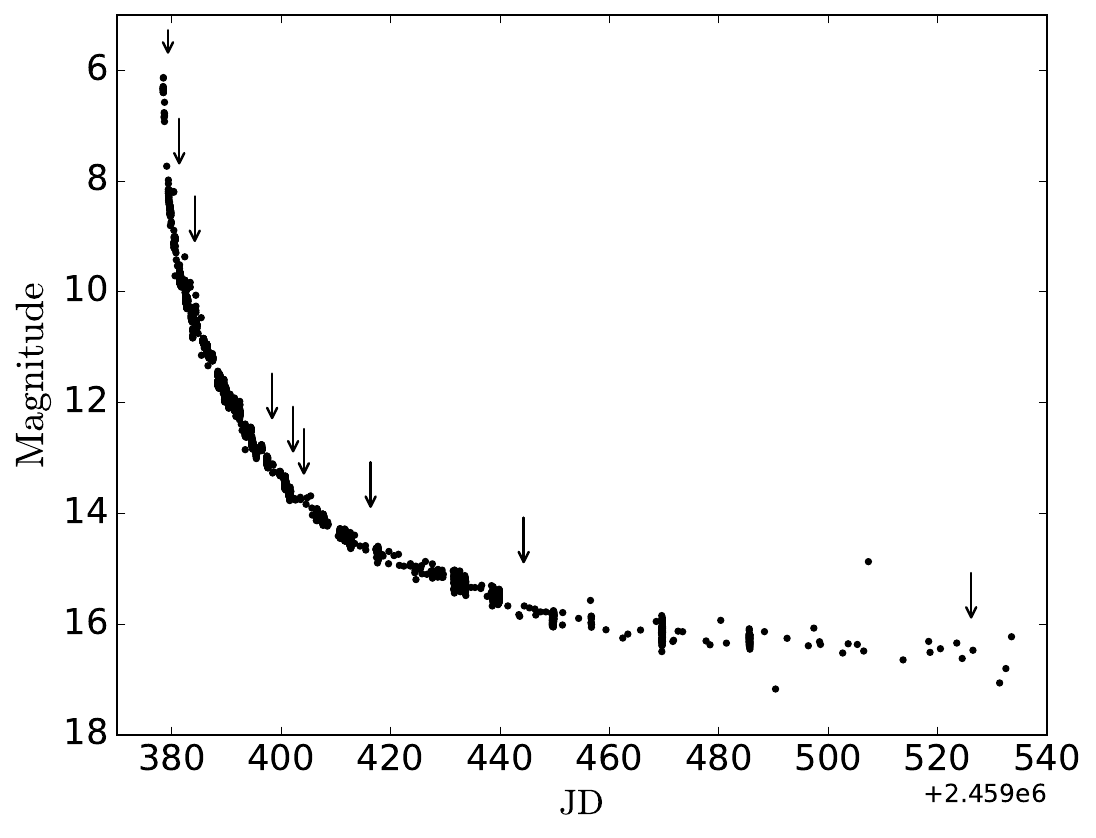}
    \caption{AAVSO V-band light curve of V1674~Her with the epochs of our spectroscopic observations marked by arrows.}
    \label{fig:AAVSO_HFOSC}
\end{figure}

Spectroscopic observations of Nova V1674~Her were obtained using the HFOSC (Himalayan Faint Object Spectrograph Camera) instrument mounted on the 2-m Himalayan Chandra Telescope (HCT) located at the Indian Astronomical Observatory (IAO), Hanle, India. The observations started on 2021 June 13 (JD 2459379.298) $\sim1$ day after the discovery (2021 June 12.537 UT \& JD 2459378.037) and continued till 2021 November 7 ($\sim148$ days). The optical spectra were obtained using Grism 7 (R $\sim1000$) and Grism 8 (R $\sim1200$) available with the HFOSC. Grisms 7 and 8 cover a wavelength range of 3800 - 8000 \AA\ and 5800 - 9200 \AA, respectively. The log of spectroscopic observations is given in Table \ref{tab:hct_log}. The epochs corresponding to our spectroscopic observations are marked on the AAVSO V-band light curve in Figure \ref{fig:AAVSO_HFOSC}. From the AAVSO database \citep{Kafka2021}, V1674~Her reached peak brightness V = 6.14 on 2021 June 12.96 UT (JD 2459378.46) \citep{2021ApJ...922L..10W}. We adopt this peak brightness time as \emph{t$_{0}$}.

All spectral data was processed using the standard task in Image Reduction and Analysis Facility (IRAF\footnote{IRAF is distributed by the National Optical Astronomy Observatory, which is operated by the Association of Universities for Research in Astronomy, Inc., under cooperative agreement with the National Science Foundation.}). The spectra were bias-corrected and flat-fielded, and the one-dimensional spectra was extracted using the optimal extraction method. Cosmic ray correction was also done. For the wavelength calibration of the extracted spectra, lamp spectra of FeNe and FeAr was used for the red and blue regions, respectively. Spectroscopic standards obtained during the same night were used for instrumental response correction. On the nights for which standard star spectra were not available, standard star data of the nearby nights was used. Then, the gr7 (blue) and the gr8 (red) spectra were combined to get the final spectrum. All the spectra were dereddened using E(B-V) = 0.55 given by \citet{2021ATel14704....1M}. The emission-line flux ratios relative to H$\beta$, measured from these dereddened spectra, are reported in Table \ref{tab:flux_ratio}.

\begin{table}
\centering
	\caption{Observational log of spectroscopic data obtained for \src}
        \label{tab:hct_log}
	\resizebox{1\hsize}{!}{\begin{tabular}{ccccccc}
			\hline
			\hline
			&  & & Exposure &   \\
			Date&JD &\emph{t-t$_{0}$}  & time & Grism \\
			dd/mm/yyyy & &(days) & (s) & & \\
			\hline	
		13-06-2021& 2459379.298 & 0.84 & 60, 60& Gr7, Gr8\\[0.25ex] 
            15-06-2021& 2459381.352& 2.89 & 60, 60& Gr7, Gr8 \\[0.25ex] 
            18-06-2021& 2459384.194& 5.73 &300, 300& Gr7, Gr8 \\[0.25ex] 
            02-07-2021& 2459398.333& 19.87 &300, 300& Gr7, Gr8 \\[0.25ex]  
            06-07-2021& 2459402.169& 23.71 &900, 900& Gr7, Gr8 \\[0.25ex]  
            08-07-2021& 2459404.137& 25.68 & 900, 900& Gr7, Gr8\\[0.25ex]  
            20-07-2021& 2459416.341& 37.88 & 900, 900 & Gr7, Gr8 \\[0.25ex]  
            17-08-2021& 2459444.324& 65.86 & 2700, 2700 & Gr7, Gr8 \\[0.25ex]  
            07-11-2021& 2459526.120& 147.66 & 1800 & Gr7 \\[0.25ex]			

			\hline
	\end{tabular}}
\end{table}

\section{Orbital \& Spin Periodicity} \label{Period}
\begin{figure}
    \centering
    \includegraphics[width=\columnwidth]{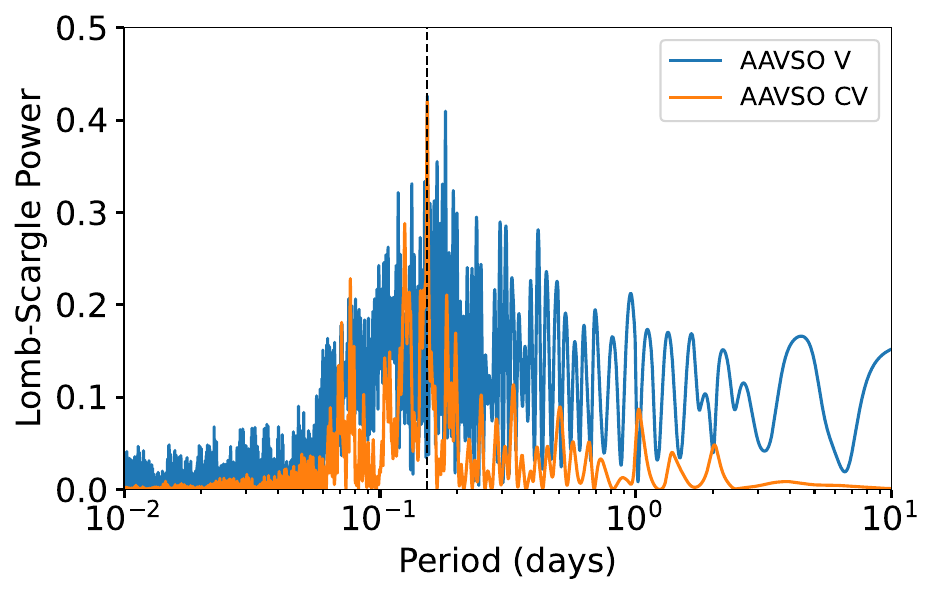}
    \caption{LS periodogram of \src\ for searching the orbital period in AAVSO V and CV bands. Both V and CV bands show a clear peak at $\sim0.153$ days, as shown by the dashed line.}
    \label{fig:lsp_aavso}
\end{figure}

Lomb-Scargle (LS) technique \citep{lomb1976Ap&SS..39..447L, scargle1982ApJ...263..835S, vdp2018ApJS..236...16V} is used for calculating the orbital and spin periodicity of \src\ using the V- and CV-band magnitude available from the AAVSO archive. During the nova eruption, it is essential first to detrend the data to find the periodicities. The detrending was performed by fitting a univariate spline with AAVSO V- and CV-band light curves. 
The detrended data covering JD 2459430 to JD 2459440 (2021 August 3.5 - 2021 August 13.5) in the V and CV band are used to obtain the LS periodogram as in Figure \ref{fig:lsp_aavso}, and from the highest peak we determined an orbital period of 0.15307 days using the CV band and 0.15321 days using the V band. These results are consistent with the findings of \citet{2021ATel14835....1S}, \citet{2022ApJ...940L..56P} and \citet{2024BAAA...65...60L}, all of whom detected a $\sim0.153$d orbital period in optical data. \citet{2024BAAA...65...60L} additionally reported a longer period of $\sim0.537$d in their TESS observations; however, no significant peak around this period is seen in our AAVSO periodogram.

\begin{figure}
    \centering
    \includegraphics[width=\columnwidth]{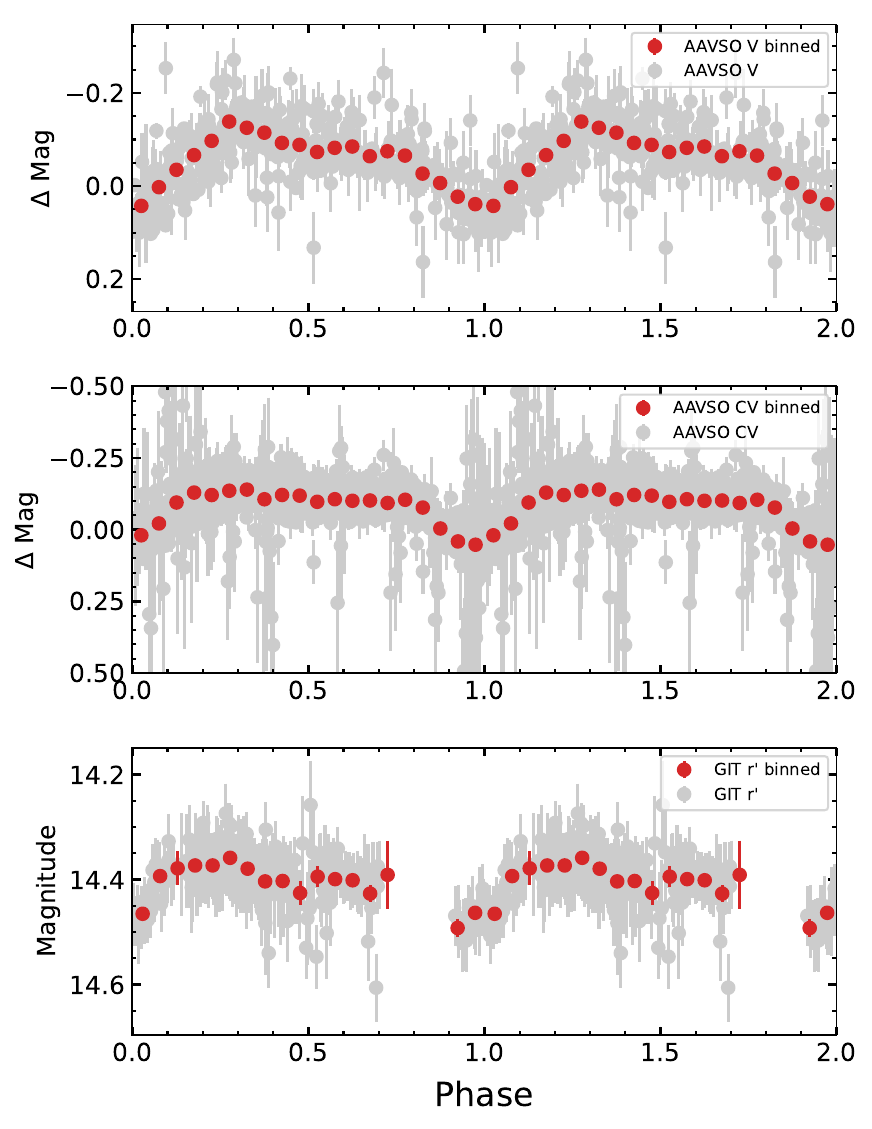}
    \caption{Phase folded light curve of \src\ in AAVSO V, CV and GIT r$^\prime$ bands. The lightcurves are folded at the period seen in Figure \ref{fig:lsp_aavso}. The GIT lightcurves also show the orbital variation but due to a lack of sufficient sampling, the sole GIT observations are unable to constrain the orbital period.}
    \label{fig:phase_folded}
\end{figure}

\begin{table*}
\centering
\caption{Line identification and observed fluxes relative to H$\beta$, derived from dereddened spectra.}
\label{tab:flux_ratio}	
	\begin{tabular}{cccccccccc}
		\hline
		\hline
            $\lambda$  & & & & $F_{\lambda}/F_{\text{H}\beta}$ & & & & \\
            \cline{2-10}\\
            (\AA) & day 0.84 & day 2.89 & day 5.73 & day 19.87 & day 23.71 & day 25.68 & day 37.88 & day 65.86 & day 147.66 \\
		\hline
			
		[Ne III 3869] + H$\zeta$ (3889) & 0.34 & 0.34 & 0.51 & 1.51 & 2.22 & 2.53 & 5.24 & 1.25 & -- \\ 
            H$\epsilon$ (3970) + [Ne III 3968] & 0.19 & 0.13 & 0.17 & 0.30 & 0.45 & 0.57 & 1.47 & 3.59 &	-- \\ 
            H $\delta$ (4102) & 0.58 & 0.53 & 0.43 & 0.65 & 0.62 & 0.65 & 0.59 & 0.82 & --noisy \\ 
            H$\gamma$ (4340) + [O III] 4363* & 0.62 & 0.52 & 0.53 & 0.68 & 0.78* & 0.79* & 1.35* & 1.92* & 0.97 \\  
            He II 4686 & --	& 0.48 & 0.77 & 1.87 & 2.32 & 2.41 & 3.60 & 4.12 & 2.01\\  
            H$\beta$ (4861) & 1 & 1 & 1 & 1 & 1 & 1 & 1 & 1 & 1\\  
            He I 5876 + Fe II* & 0.23* & 0.16* & 0.34 & 0.24 & 0.17 & 0.18 & -- & -- & --\\  
           H$\alpha$ (6563) & 2.47 & 3.31 & 4.22 & 4.16 & 3.39 & 3.57 & 3.36 & 2.65 & 1.30\\  
            He I 7065 & 0.19 & 0.13 & 0.27 & 0.20 & 0.16 & 0.16 & 0.16 & -- & --\\ 
            OI 7774 & 0.31 & 0.14 & 0.09 & -- & -- & -- & -- & -- & --\\  
            OI 8446 & 0.36 & 0.52 & 0.69 & 0.25 & 0.15 & 0.18 & 0.11 & -- & \\  
            Flux H$\beta$ (ergs/cm\textsuperscript{2}/sec)	& 8.09E-10 & 7.44E-11 & 7.07E-11 & 6.41E-12 & 1.02E-11 & 5.25E-12 & 1.79E-12 &1.86E-13 & 2.94E-14\\ 
			\hline
	\end{tabular}
\end{table*}

The light curves are phase-folded at the period of $\sim0.153$ days and plotted in Figure \ref{fig:phase_folded}, which includes the phase-folded light curves of AAVSO V and CV, as well as GIT r$^\prime$ bands. For the low-cadence GIT data, we could not determine the orbital period because it is not well-sampled enough. This can be seen in the phase-folded light curve of the GIT r$^\prime$ band, where gaps indicate insufficient data coverage. Visual inspection of the high-cadence data (see Figure \ref{git_lc}) shows sudden rises and falls but no noticeable periodicity. To further investigate this, we used the LS periodogram in a frequency range of 100 to 200 cycles/day. The periodogram showed a peak at around 500 s ($\sim172$ cycles/day), but the LS power of the highest peak was 0.04 for the r$^\prime$ filter and 0.036 for the g$^\prime$ filter, and the false alarm probability (FAP) value was close to 1.

\begin{figure*}
    \centering
    \includegraphics[width=2\columnwidth]{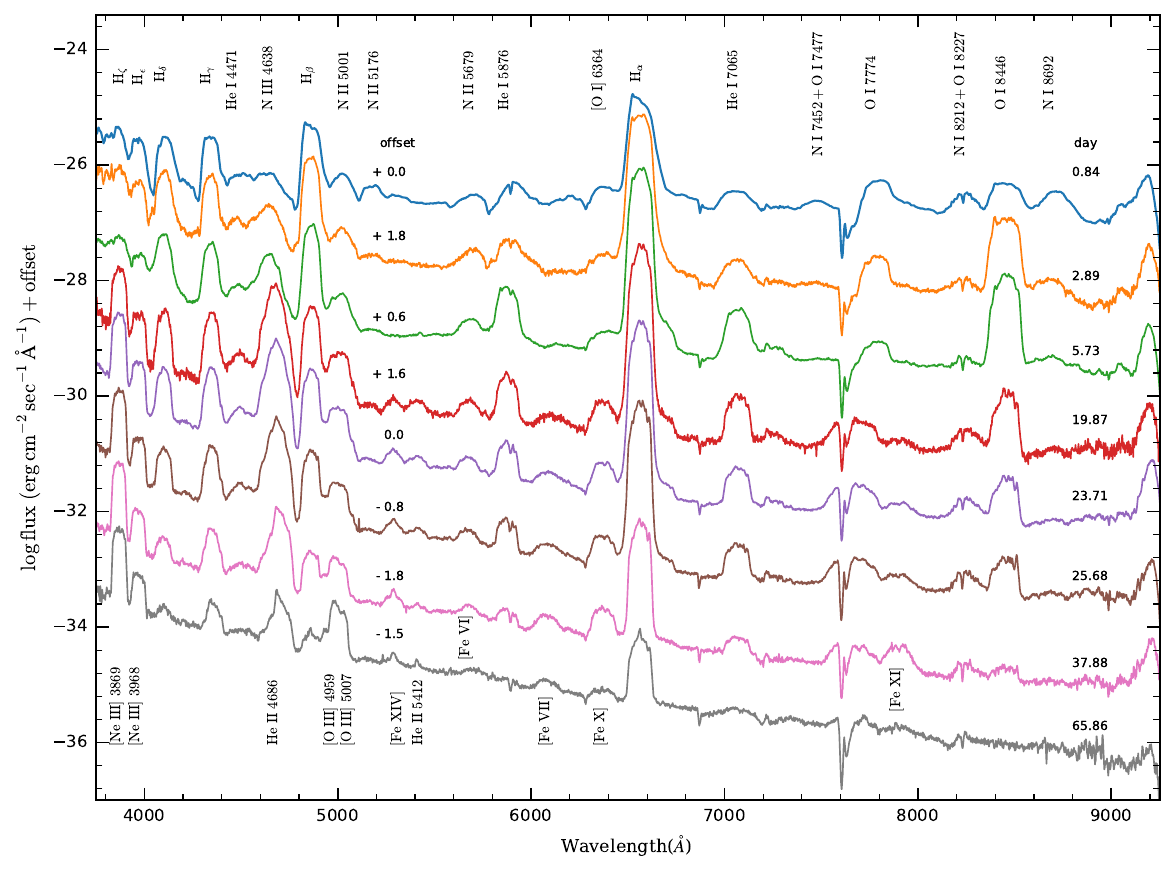}
    \caption{Spectral evolution of V1674~Her across eight epochs. Each epoch is labelled on the right side of the plot. The spectra have been given an offset for clarity. On the y-axis, the logarithm of flux is displayed, while the x-axis represents the wavelength. The logarithmic scale has been used to emphasise the visibility of weaker features.}
    \label{fig:spec_evol}
\end{figure*}

To place an upper limit on the detection of the spin modulation, we simulated 500 light curves with similar noise statistics to the current observations and injected sinusoidal signals of varying amplitudes at a fixed frequency corresponding to a 501.48 s period calculated by \citet{2022ApJ...940L..56P}. The LS periodogram was computed for each combined signal, for the frequency range between 10 to 400 cycles/day and for 10000 bins. We computed the false alarm probability (FAP) for the highest peak in the periodogram, and we noted the amplitudes for which the FAP is lower than 1\% (which indicates that at 3$\sigma$ level, we do not detect the pulsations at these amplitudes). For the observations with g$^\prime$ filter, the upper limit on the amplitude is 0.065, and for observations with r$^\prime$ filter, the upper limit is 0.04.
GIT high-cadence observations were taken around day 40 after the maximum, and the non-detection of the spin period suggests that any spin modulations in the optical data are smaller than the upper limit on amplitudes detected in our simulations. This finding aligns with the results of \citet{2022ApJ...940L..56P}, who reported variations of around 0.01 magnitudes on day 12, increasing to about 0.09 magnitudes by day 350.

\begin{figure*}
   \centering
    \includegraphics[width=2\columnwidth]{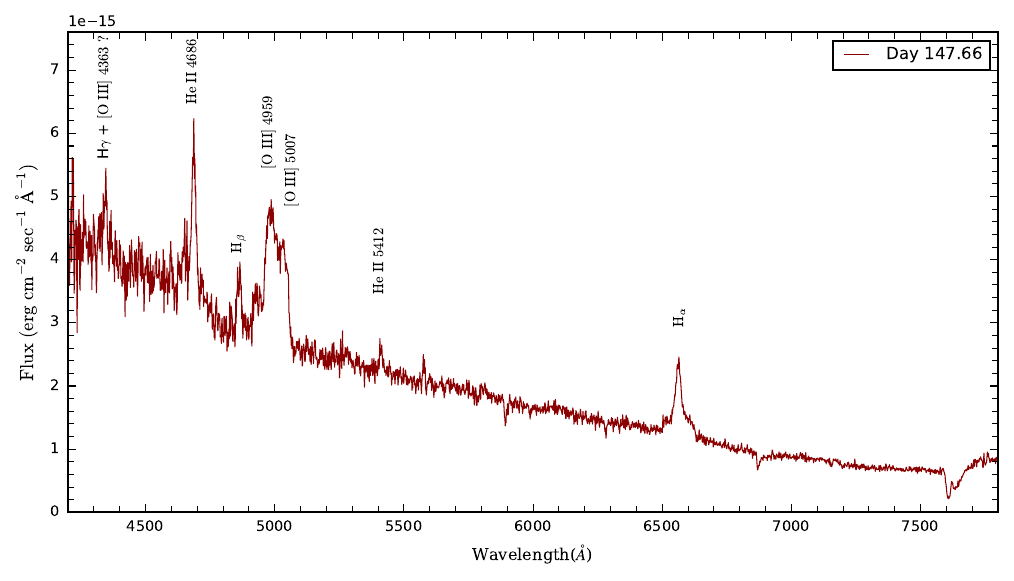}
   \caption{Spectrum of V1674~Her on day +147.66. The nebular lines are still visible and are very strong. He II 4686 \AA\ line is present and is stronger in strength compared to the Balmer lines. The presence of He II 4686 \AA\ line and the rising continuum in the blue indicate an accretion-dominated spectrum.}
   \label{fig:7Nov_2021_spectra}
\end{figure*}

\begin{figure}
    \centering
    \includegraphics[width=\columnwidth]{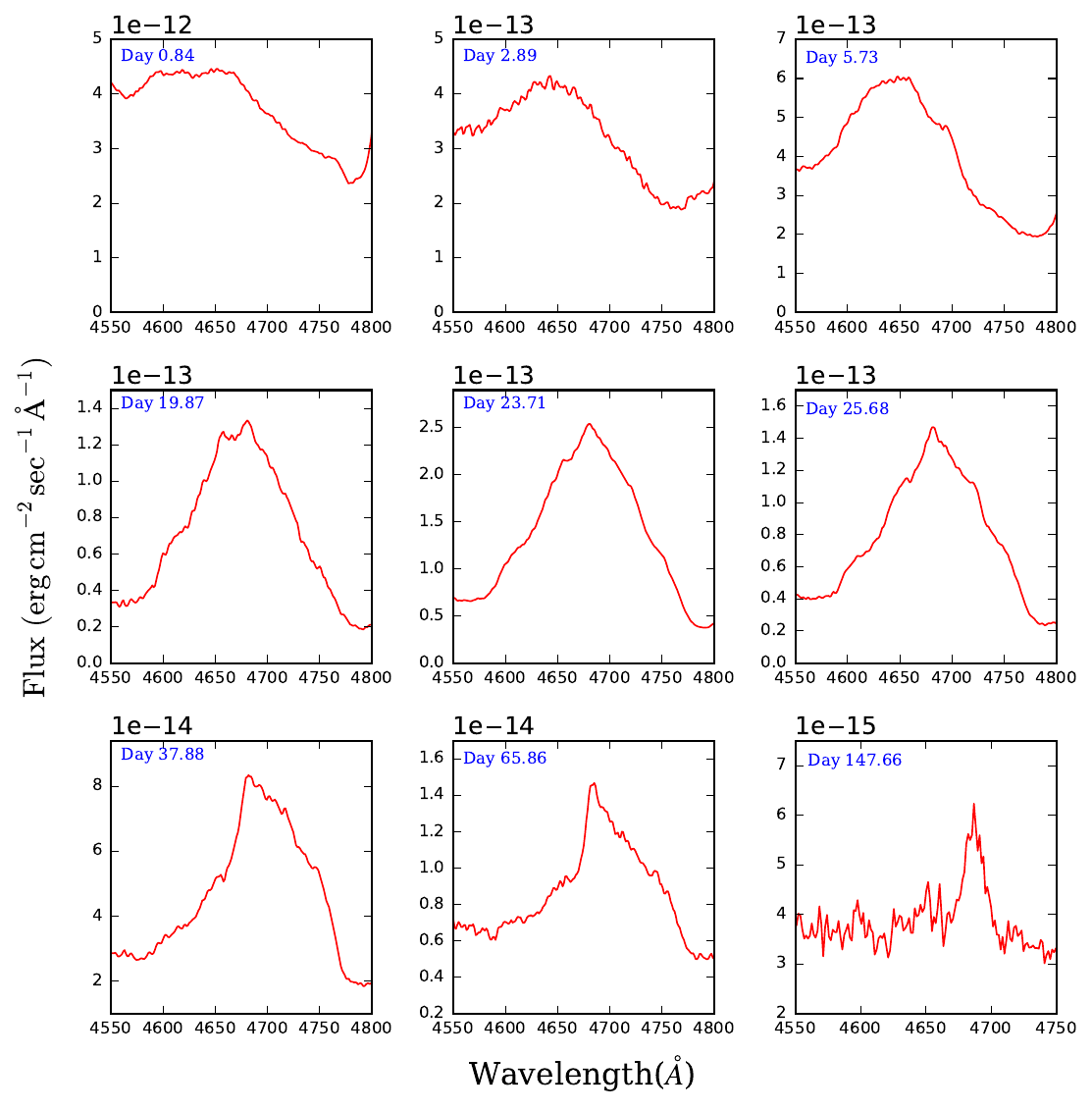}
    \caption{Evolution of the He II 4686 \AA\ line profile across nine epochs. Initially, on day 0.84, the contribution arises predominantly from N III 4638 \AA, potentially accompanied by the Fe II (37) multiplet at 4629 \AA. By day 19.87, He II 4686 \AA\ is distinctly present. The appearance of [Ne III] and [Ne V] lines on day 19.87 (see Section \ref{Neon}) presents a compelling case for the presence of [Ne IV] 4721 \AA\ line, albeit possibly blended within the broad He II 4686 \AA\ feature. This blending might be one of the reasons for the asymmetry observed in the shape of the He II 4686 line profile on day 37.88. Furthermore, on day 147.66, a considerable decrease in the full width at half maximum (FWHM) is observed, indicating a potential contribution from a different region or due to wind from the accretion disk.}
    \label{fig:He_II}
\end{figure}

\begin{figure}
    \centering
    \includegraphics[width=\columnwidth]{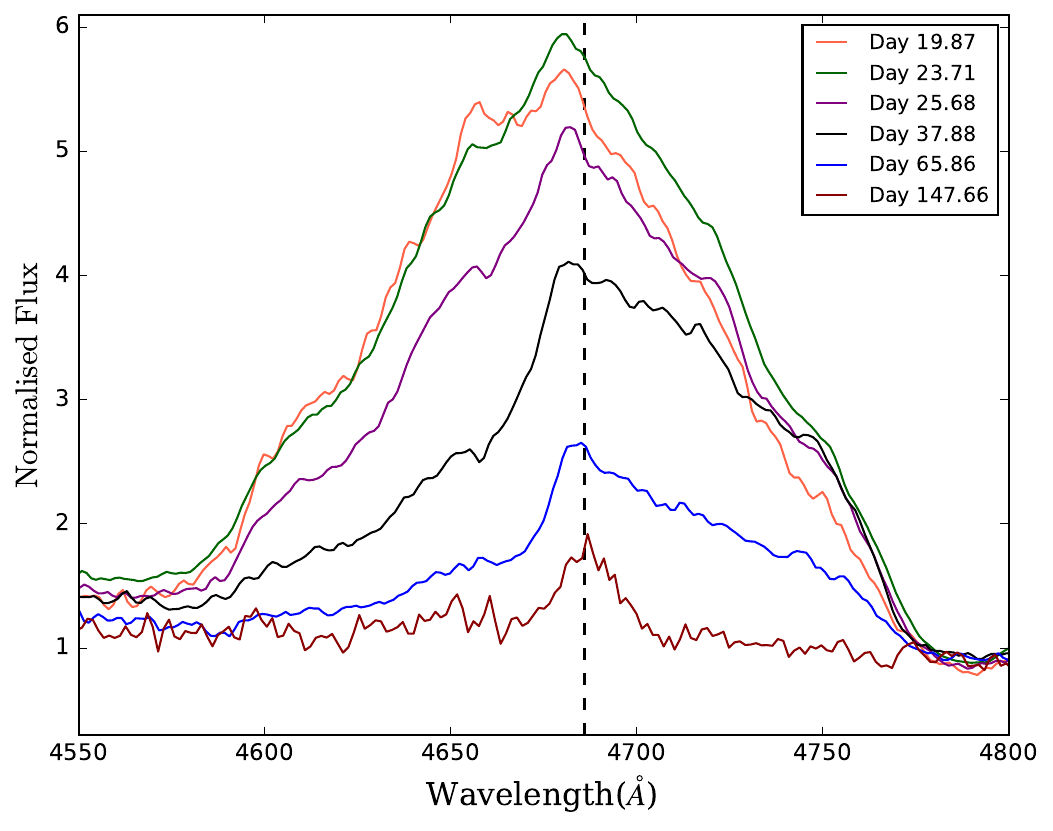}    
    \caption{Emission line profiles of He II 4686 \AA. Dotted line represents the rest wavelength of He II 4686 \AA. For each epoch, the flux has been normalised to the local continuum. He II 4686 \AA\ is clearly observed on day 19.87 and is present throughout till day 147.66.}
    \label{fig:HeII_Variation}
\end{figure}

\begin{figure}
   \centering
    \includegraphics[width=\columnwidth]{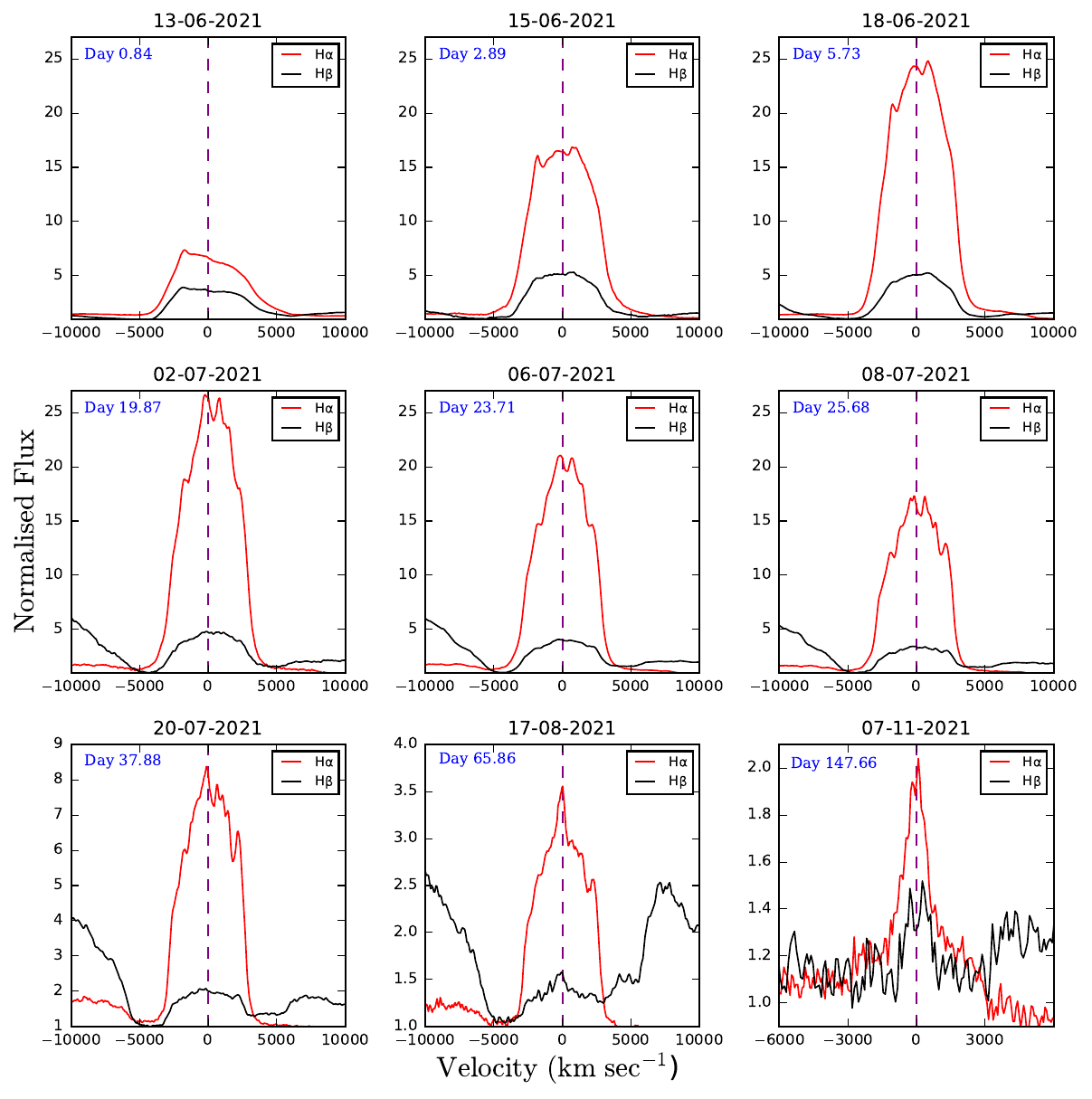}
   \caption{Line profiles of H$\alpha$ and H$\beta$ for nine epochs. For each epoch, the H$\alpha$ and H$\beta$ flux values have been normalised to their respective local continuum. A common y-axis scale is adopted for the early epochs, while a reduced scale is used for the later epochs, where the line fluxes are significantly weaker. The lines display corrugated profiles with generally similar shapes. On day 0.84, asymmetry is observed, characterised by a prominent blue spike atop a flat profile, which transitions to symmetry over time. The FWHM velocities remain relatively stable until day 65.86; however, on day 147.66, FWHM values differ significantly from the previous ones. These notably different FWHM values suggest a shift in contribution, potentially originating from a different region or due to wind from the accretion disk.}
   \label{fig:H_alpha_beta_allday}
\end{figure}

\begin{figure}
    \centering
    \includegraphics[width=\columnwidth]{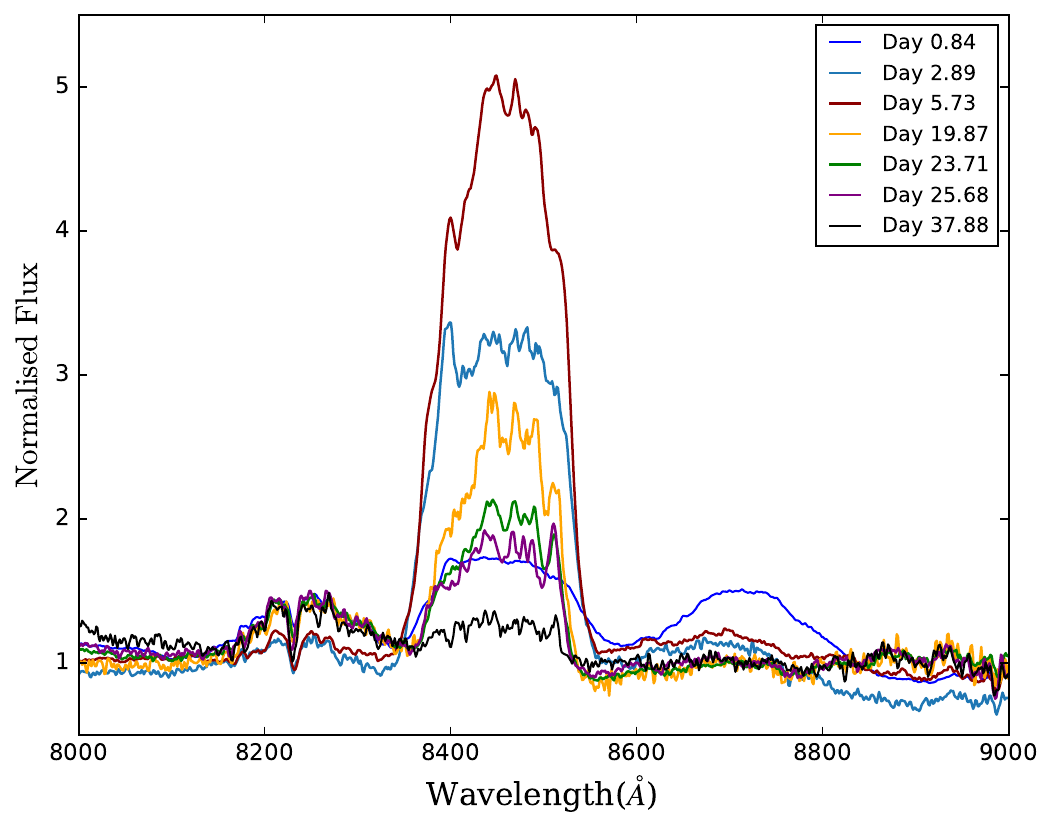}
    \caption{Evolution of the O I 8446 \AA\ line. For each epoch, the flux has been normalised to the local continuum. The line reaches a peak in strength by day 5.73 and subsequently declines in later epochs.}
    \label{fig:OI8446}
\end{figure}

\begin{figure*}
    \centering
    \includegraphics[width=2\columnwidth]{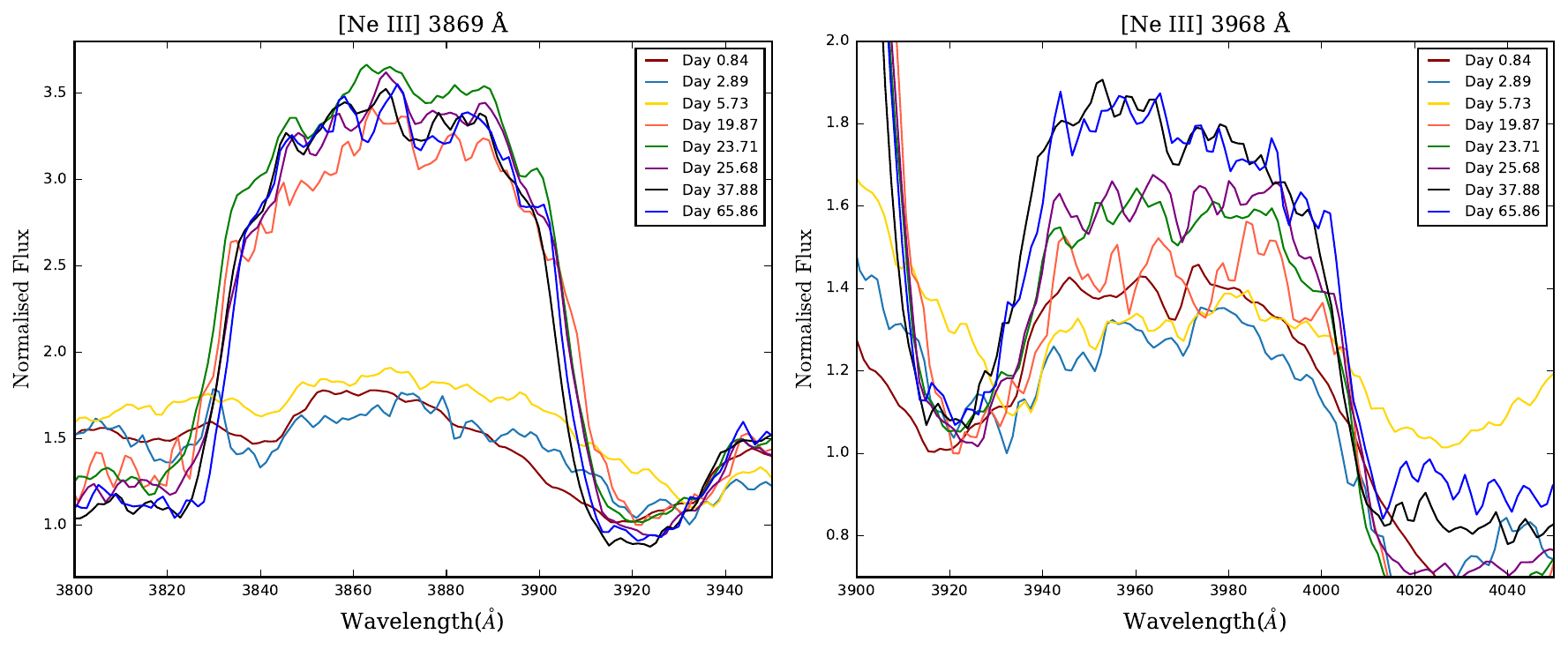}
    \caption{Evolution of the [Ne III] 3869 \AA\ (left) \& [Ne III] 3968 \AA\ (right) emission lines. For each epoch, the [Ne III] 3869 \AA\ \& [Ne III] 3968 \AA\ flux values have been normalised to their respective local continuum. In the initial spectra (day 0.84, 2.89 \& 5.73) obtained shortly after the nova eruption, prominent hydrogen lines H$\zeta$ \& H$\epsilon$ are observed. However, as the eruption progressed and transitioned into the SSS phase on day 18.9 \citep{2021ATel14747....1P}, the [Ne III] 3869 \AA\ \& [Ne III] 3968 \AA\ lines are observed in the spectra starting from day 19.87. These neon lines remained present in subsequent spectra until the epoch of day 65.86. They are also present on day 147.66 (see Section \ref{Later_spectra}).}
    \label{fig:[NeIII]_Variation}
\end{figure*}

\begin{figure}
    \centering
    \includegraphics[width=\columnwidth]{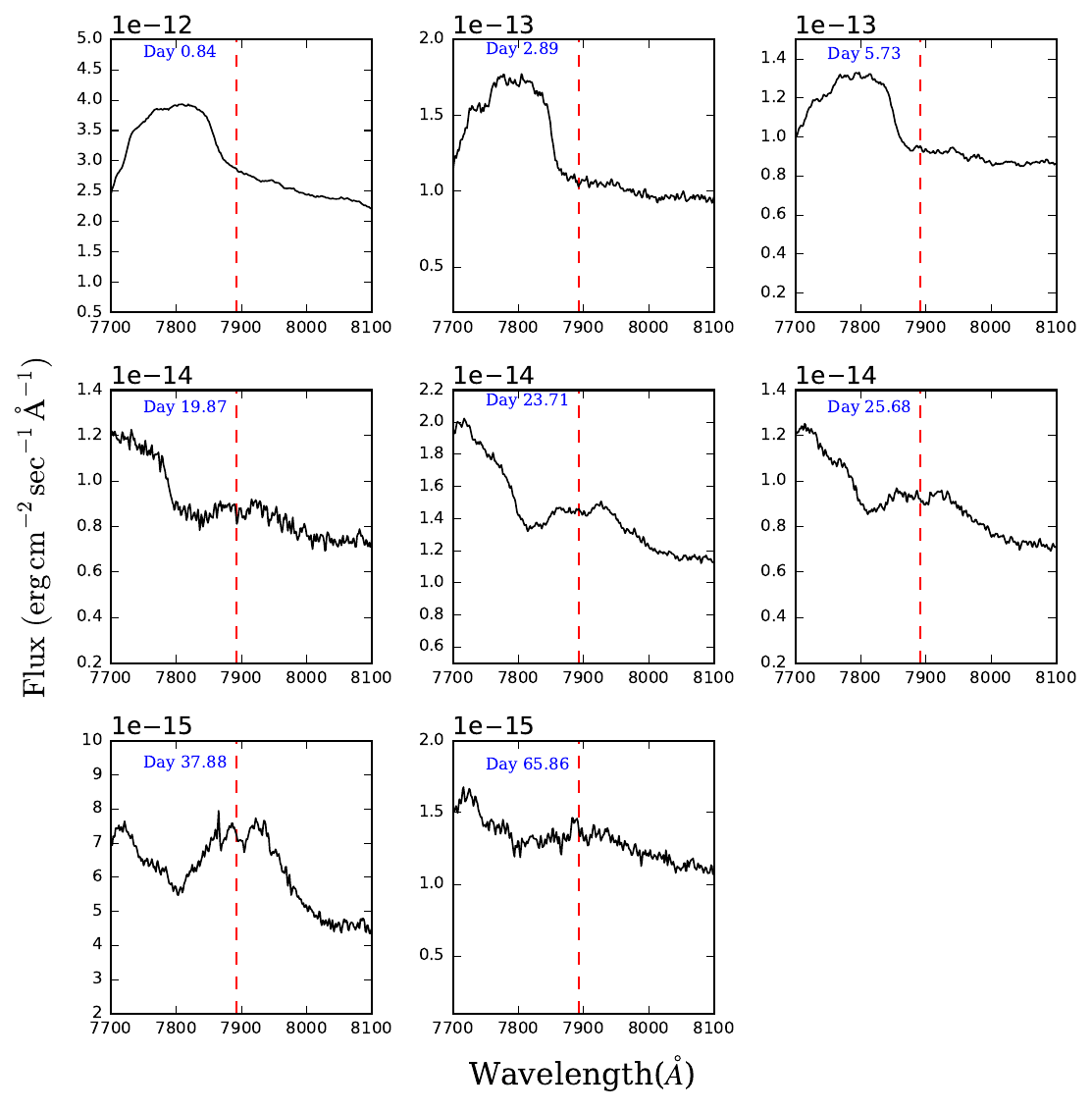}
    \caption{Evolution of [Fe XI] 7892 \AA. Dotted line represents the rest wavelength of [Fe XI] 7892 \AA. In the early spectra (day 0.84, 2.89, \& 5.73), the O I 7774 \AA\ feature is visible. A weak emission line of [Fe XI] 7892 \AA\ is observed on day 23.71, which becomes more prominent on day 37.88. However, [Fe XI] is absent on day 65.86.}
    \label{fig:[Fe XI] 7892}
\end{figure}

\begin{figure}
    \centering
    \includegraphics[width=\columnwidth]{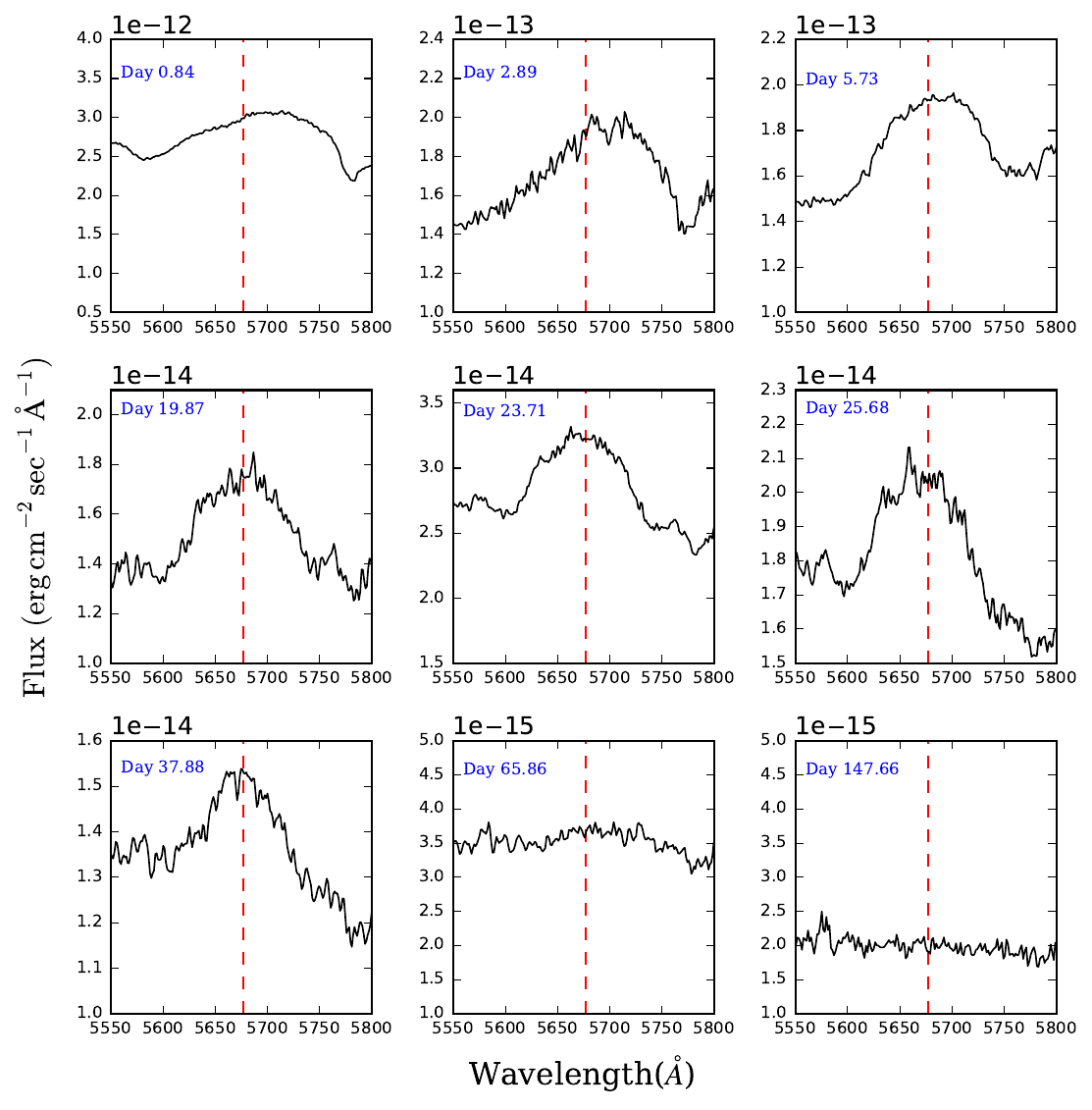}
    \caption{Evolution of [Fe VI] 5677 \AA\ emission line. Dotted line represents the rest wavelength of [Fe VI] 5677 \AA. Initially, on day 0.84, the observed emission feature is attributed to N II 5679 \AA. Subsequently, on day 19.87, [Fe VI] 5677 \AA\ becomes distinctly present and persists until day 37.88. However, [Fe VI] is substantially weak on day 65.86 and eventually disappears by day 147.66.}
    \label{fig:[FeVI_5677]}
\end{figure}

\section{Spectral Evolution} \label{SpectralEvolution}
The spectral evolution of V1674~Her over eight epochs is shown in Figure \ref{fig:spec_evol}. Initially, the spectrum is dominated by low ionisation lines of helium, hydrogen, oxygen and nitrogen. Possible contributions from Fe II transitions may be present in the earliest epoch, but these features are strongly blended with other broad emission lines, making individual Fe II lines difficult to identify, so they are not explicitly labelled in Figure \ref{fig:spec_evol}. Fe II lines disappear very soon after the nova eruption. As the ejecta expands and the layers closer to the central WD are revealed, the ionisation levels increase. By day 19.87 after eruption, coronal (e.g. [Fe X] [Ionisation Potential (IP) = 235.04 eV], [Fe XIV] [IP = 361 eV]) and high ionisation lines of iron (e.g. [Fe VI] [IP = 75 eV], [Fe VII] [IP = 99 eV]), along with nebular lines (e.g. [O III] [IP = 35.116 eV], [Ne III] [IP = 40.962 eV]) show up in the spectra. All these lines increase in strength over time before weakening. The spectrum on day 147.66 (see Figure \ref{fig:7Nov_2021_spectra}) shows that the coronal and high ionisation lines have eventually disappeared. However, the nebular lines are still present. Although He II 4686 \AA\ is present, there has been a subsequent diminishing of the full width at half maximum (FWHM) of He II 4686 \AA\ (see Figures \ref{fig:He_II} \& \ref{fig:HeII_Variation}). Furthermore, the spectrum shows a rising continuum in the blue, and the presence of the He II 4686 \AA\ line indicates an accretion-dominated spectrum. The detailed evolution of the optical spectra of V1674~Her is given in the subsequent sub-sections.

\subsection{Early Spectra: day 1 to 6} \label{EarlySpectra}
\textbf{Day 0.84:} The early spectrum is dominated by lines of helium, oxygen and nitrogen. Typical ejecta velocities, calculated using FWHM, are approximately 6300 km/s for helium, 5800 km/s for oxygen, and 5150 km/s for nitrogen lines. Hydrogen Balmer lines are also present, with FWHM velocities of 5830 km/s for H$\alpha$, 5800 km/s for H$\beta$, and 5780 km/s for H$\gamma$. The Balmer emission lines also show very broad wings; the full width at zero intensity (FWZI) velocity, estimated from the velocity extent where the emission merges with the local continuum, reaches $\sim$10,600 km/s for H$\alpha$ and $\sim$9,300 km/s for H$\beta$. H$\alpha$ and H$\beta$ emission lines exhibit a blue shoulder atop their flat top profiles. The Balmer-line peaks show a corrugated structure, indicating multiple substructures within the ejecta (see Figure \ref{fig:H_alpha_beta_allday}). H$\gamma$ exhibits a structural difference from H$\alpha$ and H$\beta$, possibly due to the contribution of other lines blended with the H$\gamma$ feature. He I 7065 \AA\ is clearly present, while He I 5876 \AA\ is possibly blended with another line. He I 5016 \AA\ \& 5048 \AA\ may also be present but could be blended with N II 5001 \AA\ line. N I 8692 \AA\ and N II 5679 \AA\ are clearly present. The combination of O I 8227 \AA\ \& N I 8212 \AA\ and O I 7477 \AA\ \& N I 7452 \AA\ is also observed. Interstellar Na I (D) absorption lines are also identifiable.

On the left side of H$\beta$, two distinct broad features can be observed centred at 4622 \AA\ and 4507 \AA. These features are likely due to the blending of several emission lines. The feature at 4622 \AA\ is N III 4638 \AA, with a possible contribution from Fe II (37) multiplet at 4629 \AA. The 4507 \AA\ feature is likely a blend of N III 4517 \AA, He I 4471 \AA, Fe II (37) multiplet at 4491 \AA\ \& 4515 \AA, and Fe II (38) multiplet at 4508 \AA, 4523 \AA\ \& 4549 \AA. Moving rightward from H$\beta$, a feature centred at 5014 \AA, is likely attributable to N II 5001 \AA, with possible contributions from He I 5016 \AA\ \& 5048 \AA, and Fe II (42) multiplet at 5018 \AA. However, the presence of Fe II is not definitive. The feature centred at 5179 \AA\ is tentatively assigned to a blend of N II 5176 \AA\ \& 5180 \AA, and Fe II (42) multiplet at 5168 \AA, along with Fe II (49) multiplet at 5199 \AA. The feature to the right of $\lambda$ $\sim5179$ possibly originates from a combination of Fe II (49) multiplet at 5276 \AA\ \& 5317 \AA\ and Fe II (41) multiplet at 5284 \AA. Additionally, the oxygen lines O I 8446 \AA\ and O I 7774 \AA\ are also present. O I 8446 \AA\ also has a corrugated structure with multiple sub-peaks, indicating many velocity components are being observed (see Figure \ref{fig:OI8446}). The strength of the O I 8446 \AA\ is slightly more than the O I 7774 \AA\ emission line. There is an emission line left to the H$\alpha$, possibly due to the [O I] 6364 \AA.

The early spectra of V1674~Her share similarities with the spectra of V2491 Cyg (Nova Cyg 2008 N.2) given by \cite{2011NewA...16..209M}, particularly in the structure of He I 5876 \AA\ and $\lambda$ $\sim8231.482$ \AA\ (N I + O I) feature. V2491 Cyg, discovered by K. Nishiyama and F. Kabashima at $\sim7.7$ mag \citep{2008IAUC.8934....1N}, was classified as a He/N class nova. It displayed a $t_2$ time of 4.8 days, with a secondary maximum also observed. It had a E(B-V) value of 0.24, and the distance was 14kpc. Multiple absorption features, such as Na I 8191 \AA\ and Na I (D) lines, among others, are also evident in the spectra of V2491 Cyg.

On \textbf{day 2.89}, the H$\alpha$ line profile shows a red peak on top of a flat profile (see Figure \ref{fig:H_alpha_beta_allday}). The same is the case for H$\beta$ and H$\gamma$. Additionally, there seems to be a contribution on the red side tail of the H$\alpha$ feature, which could possibly be due to He I 6678 \AA\ line. Overall, the flux values have decreased in comparison to day 0.84. The O I 8446 \AA\ line has increased in strength compared to the O I 7774 \AA\ line. The O I 8446/7774 ratio = 3.64, indicating that neutral oxygen is predominantly excited by Lymann-beta fluorescence (see Section \ref{Oxygen}).

On \textbf{day 5.73}, He II 4686 \AA\ seems to be present on the right edge of the feature centred at 4637 \AA. However, the broad N III 4638 \AA\ line makes it difficult to confirm the presence of the He II 4686 \AA\ line (see Figure \ref{fig:He_II}). H$\alpha$ and H$\beta$ continue to exhibit a red peak in their line profiles (see Figure \ref{fig:H_alpha_beta_allday}). H$\gamma$ also shows enhanced emission towards the red side. In contrast, H$\delta$ has a boxy peak without a pronounced red component. Furthermore, the He I 6678 \AA\ line has become evident on the tail of H$\alpha$. Both He I 5876 \AA\ and 7065 \AA\ have increased in strength relative to the continuum compared to day 2.89 spectra. The strength of O I 8446 \AA\ has increased further (see Figure \ref{fig:OI8446}), and the O I 8446 / 7774 ratio = 7.93.

\subsection{Later Spectra: day 20 to 148}\label{Later_spectra}
\citet{2021ATel14747....1P} reported that V1674~Her entered the SSS phase on day 18.9. The spectra obtained on \textbf{day 19.87} show the appearance of high-ionisation and coronal lines of Fe: [Fe VI] 5677 \AA\ (IP = 75 eV), [Fe VII] 6087 \AA\ (IP = 99 eV), [Fe X] 6375 \AA\ (IP = 235.04 eV) and [Fe XIV] 5303 \AA\ (IP = 361 eV). Additionally, [Ne III] 3869 \AA\ and 3968 \AA\ (IP = 40.962 eV) lines are present, with [Ne III] 3869 \AA\ being particularly prominent, indicating the presence of an ONe-type WD. Both [Ne III] 3869 \AA\ \& 3968 \AA\ show corrugated structures with multiple sub-peaks (see Figure \ref{fig:[NeIII]_Variation}). H$\zeta$ (3889 \AA) and H$\epsilon$ (3970 \AA) lines are blended with the [Ne III] 3869 \AA\ and 3968 \AA\ lines, respectively. The flux ratio of ([Ne III] 3869 \AA\ + H$\zeta$) \& ([Ne III 3968] + H$\epsilon$) feature with respect to the H$\beta$ (see Table \ref{tab:flux_ratio}) is 1.51 and 0.30, respectively, which is higher than the ratios on day 5.73, indicating that the dominant contribution is due to the Ne III lines in both features and ascertaining the presence of [Ne III] lines. Furthermore, [Ne V] (IP = 97.11 eV) 3426 \AA\ is visible, but the response of Grism 7 is not good in the region less than 3800 \AA; hence, the flux values calculated in this region are not very accurate. Nebular lines of oxygen, [O III] (IP = 35.116 eV) 4959 \AA, 5007 \AA\ are present in the spectra. The presence of these coronal, nebular and high ionisation lines indicates that the ionisation levels have increased. He II 4686 \AA\ is clearly present for the first time and shows a redder peak (see Figure \ref{fig:HeII_Variation}). Moreover, [Ne IV] (IP = 63.45 eV) 4721 \AA\ might be present but is blended with the broad He II 4686 \AA\ feature. He II 5412 \AA\ is also present. Furthermore, the strength of the O I 8446 \AA\ line has reduced and is now comparable to the He I 7065 \AA\ line. This reduction indicates a decrease in the neutral gas and highlights the increase in ionisation levels.

\textbf{Day 23.71:} In comparison to day 19.87, based on the lines present, the spectrum on day 23.71 has not changed significantly. The value of observed fluxes relative to H$\beta$ for nebular lines has increased, while the value for H$\alpha$ has decreased (see Table \ref{tab:flux_ratio}). In contrast, the H$\gamma$/H$\beta$ flux ratio has increased, indicating an additional contribution to the flux value of H$\gamma$, possibly due to another line blended with the broad H$\gamma$ feature. This additional contribution could be due to [O III] 4363 \AA\ line. Additionally, a weak emission line of [Fe XI] (IP = 261.1 eV) 7892 \AA\ is also observed (see Figure \ref{fig:[Fe XI] 7892}).

Around the same time, UV spectra from \citet{2024MNRAS.528...28B} (\textbf{day 24}) show the presence of neon emission lines, [Ne V] 1575 \AA\ and [Ne IV] 1602 \AA, and He II 1640 \AA\ and several other features. The presence of these neon lines in the UV, combined with the optical detection of prominent neon lines, further reinforces the classification of V1674~Her as a neon nova.

\textbf{Day 25.68:} No significant changes are observed in the spectra compared to day 23.71.

\textbf{Day 37.88:} All the coronal, nebular and high ionisation lines observed on day 25.68 are present. In contrast to a weak emission on day 25.86, [Fe XI] 7892 \AA\ is distinctly present in the spectra (see Figure \ref{fig:[Fe XI] 7892}). Furthermore, the intensity of the O I 8446 \AA\ line has been significantly reduced (see Figure \ref{fig:OI8446}).

On \textbf{day 38}, UV spectra from \citet{2024MNRAS.528...28B} show a decrease in the strength of He II 1640 \AA. A similar trend is observed in the optical, where He II 4686 \AA\ emission also weakens (see Figure \ref{fig:HeII_Variation}).

\textbf{Day 65.86:} The overall flux has reduced compared to day 37.88. [Fe VI] 5677 \AA\ has subsequently weakened while [Fe XI] 7892 \AA\ is absent in the spectra (see Figures \ref{fig:[FeVI_5677]} \& \ref{fig:[Fe XI] 7892}). O I 8446 \AA\ is also not present.

\textbf{Day 147.66:} Coronal and high ionisation lines of iron have eventually disappeared. H$\alpha$, H$\beta$, H$\gamma$, He II 5412 \AA, He II 4686 \AA\ and [OIII] 4959 \AA, 5007 \AA\ lines are present, as seen in Figure \ref{fig:7Nov_2021_spectra}. The [O III] 4363 \AA\ line may also be present, but it is difficult to identify as it is blended with the H$\gamma$ feature. After smoothing the spectrum using IRAF with a smoothing factor of 8, the [Ne III] 3869 \AA\ and 3968 \AA\ lines stand out and are seen clearly. But even after smoothing, it is hard to comment if [Ne V] 3426 \AA\ is present. The FWHM velocity of He II 4686 \AA\ line is $\sim1500$ km/s, much lower than the typical FWHM velocity of more than 6000 km/s for the other ‘Later spectra'. A significantly different FWHM velocity indicates that the contribution to He II 4686 \AA\ is coming from a different region, an accretion disc which is irradiated by the hot WD. Moreover, the spectrum shows a rising continuum in the blue. The presence of the He II 4686 \AA\ line and the rising continuum in blue indicate an accretion-dominated spectrum.

\begin{figure}
   \centering
    \includegraphics[width=\columnwidth]{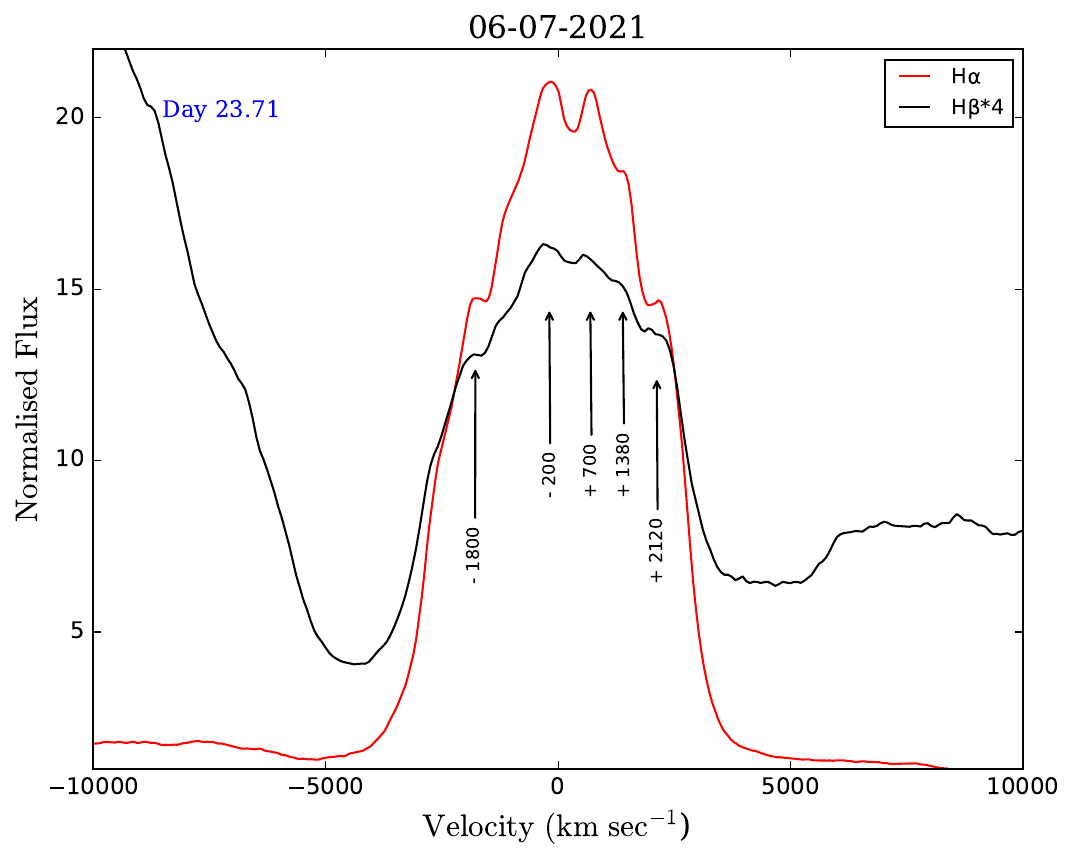}
   \caption{Velocity profiles of H$\alpha$ and H$\beta$ on day 23.71. H$\alpha$ and H$\beta$ fluxes have been normalised to their respective local continuum. Furthermore, the H$\beta$ profile has been scaled by a constant factor for clarity to highlight weaker sub-structures. Five distinct parcels of gas are evident. The similar shapes of both lines suggest that H$\alpha$ and H$\beta$ emissions originate from the same region.} \label{fig:H_alpha_beta_6thJuly}
\end{figure}

\section{Observed emission lines in V1674~Her}\label{Description}

\subsection{Hydrogen Lines}\label{Hydrogen}
Balmer lines of hydrogen are present in the spectra of V1674~Her. H$\alpha$ and H$\beta$ have similar line profiles (see Figure \ref{fig:H_alpha_beta_allday}) while the H$\gamma$ line is contaminated due to contributions from the Fe II line in the early spectra and later due to the [O III] 4363 \AA\ line (see Section \ref{Oxygen} for details). H$\delta$ has a different structure compared to the H$\alpha$, H$\beta$ and H$\gamma$. The similarity in the H$\alpha$ and H$\beta$ structure indicates that they originate from the same region. Both lines show corrugated line profiles with multiple sub-peaks. These sub-peaks suggest that there are sub-structures within the ejecta. For example, on day 23.71, five distinct parcels of gas at radial velocities -1800, -200, 700, 1380 and 2120 km/s can be seen in H$\alpha$ and H$\beta$ line profiles (see Figure \ref{fig:H_alpha_beta_6thJuly}).

\citet{1999MNRAS.307..677G} modelled optically thin emission-line profiles for several geometries of classical nova shells across different inclination angles, where an inclination of 0\degree\ indicated that the system is viewed along the polar symmetry axis. For a simple ellipsoidal shell featuring an equatorial ring along with brighter polar rings, their simulations show that, at an inclination of i = 60\degree, the resulting profiles closely resemble the H$\alpha$ and H$\beta$ line profiles observed on day 23.71. This angle of inclination is consistent with the value found by \citet{2024MNRAS.527.1405H}, who determined an inclination of i = 67 ± 1.5\degree\ using SHAPE\footnote{\url{https://wsteffen75.wixsite.com/website}} \citep{2011ITVCG..17..454S} modelling.

There is no appreciable change in the FWHM velocity of H$\alpha$ and H$\beta$ for the first eight epochs, where it remains roughly around 4500 to 5500 km/s. However, on day 147.66, the FWHM velocity of  H$\alpha$ and H$\beta$ is $\sim1500$ km/s. This significantly different FWHM velocity suggests a change in contribution, potentially originating from a different region or due to wind from the accretion disk.

\begin{figure}
    \centering
    \includegraphics[width=\columnwidth]{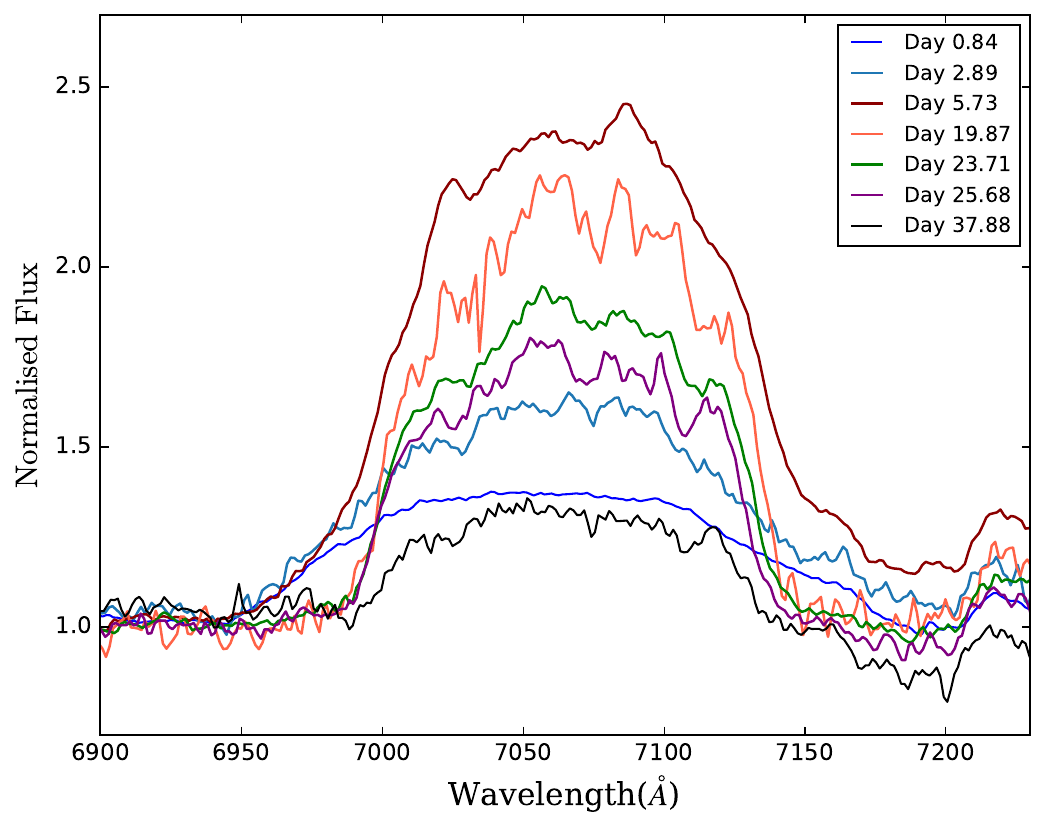}
    \caption{Evolution of the He I 7065 \AA\ line. For each epoch, the flux has been normalised to the local continuum. Initially, on day 0.84, the profile appears flat without discernible substructures. However, distinct substructures emerge in the line profile as the nova evolves. For e.g. on day 5.73, 3 substructures can be seen. By day 65.86, the He I 7065 \AA\ line has weakened significantly (see Figure \ref{fig:spec_evol}).}
    \label{fig:HeI7065}
\end{figure}

\begin{figure*}
    \centering
    \includegraphics[width=2\columnwidth]{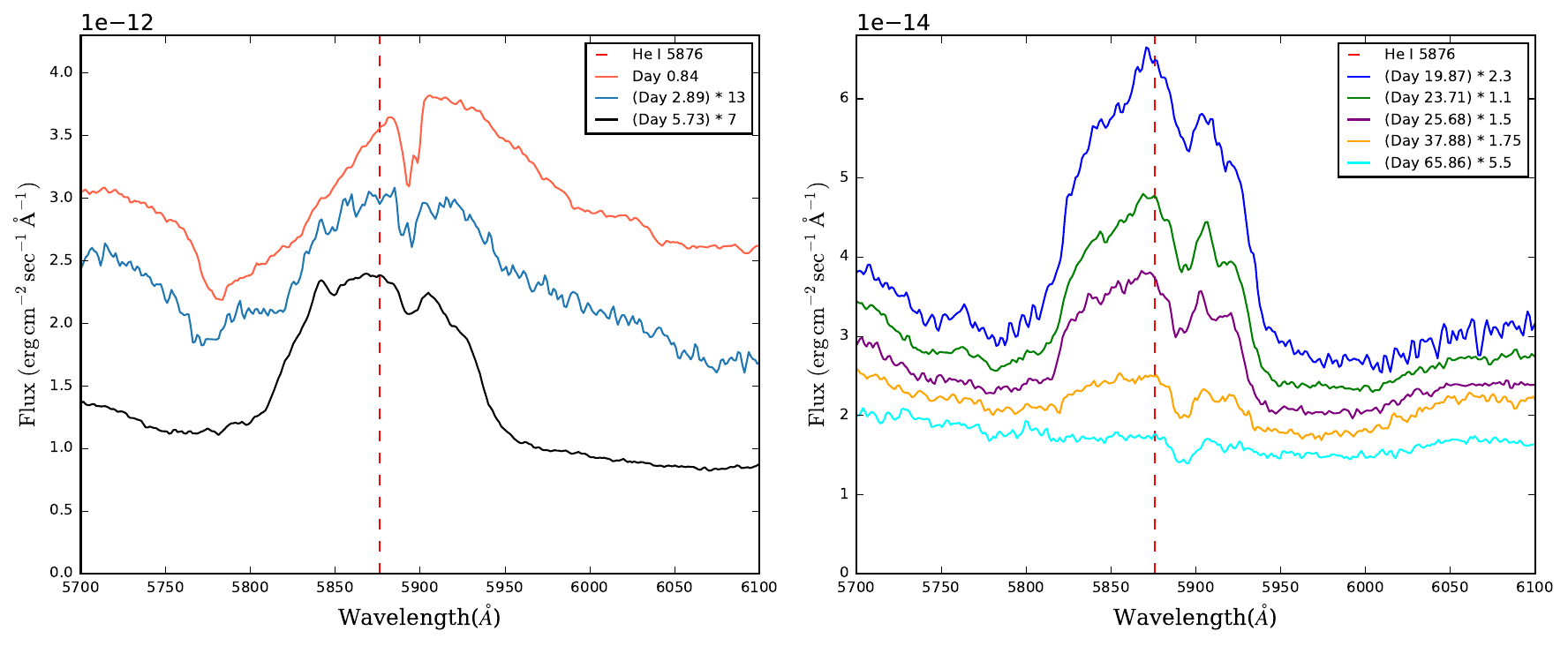}
    \caption{Evolution of the He I 5876 \AA\ line. The fluxes are scaled by a constant factor, as indicated in the legend. Dotted line represents the rest wavelength of He I 5876 \AA. By day 65.86, the He I 5876 \AA\ line has weakened significantly. Interstellar Na I (D) absorption lines are consistently observed across all epochs.}
    \label{fig:HeI5876}
\end{figure*}

\subsection{Helium Lines}
In the optical spectra of V1674~Her, both He I and He II lines appear and disappear at different times in the evolution of the nova. Initially, He I emission lines are present in the spectra. Some stand out (e.g. He I 7065 \AA, 5876 \AA), while some are blended (e.g. He I 4471 \AA, 5016 \AA, 5048 \AA) with other lines. Low ionisation He I lines initially increase in strength relative to the continuum, but as the ionisation levels further increase, the strength of the He I lines weakens, and they eventually disappear. However, He II lines appear in the spectrum as a result of the increase in the ionisation conditions.

On day 0.84, He I 7065 \AA\ is present, but the line profile is flat. Subsequently, on day 2.89, the profile becomes corrugated and increases in strength. On day 5.73, the strength of He I 7065 \AA\  peaked before gradually decreasing in the later epochs, eventually disappearing on day 147.66. The evolution of He I 7065 \AA\  is shown in Figure \ref{fig:HeI7065}.

He I 5876 \AA\ is also present on day 0.84. However, the structure of the emission line with respect to the rest wavelength of 5876 \AA\ is different on day 0.84 (see Figure \ref{fig:HeI5876}) in comparison to the later epochs. Although consistently observed until day 37.88, the line weakens significantly on day 65.86 before disappearing altogether. Interstellar Na I (D) absorption lines are also present.

He I 6678 \AA\ is possibly present on the tail of H$\alpha$ on day 2.89 (see Figure \ref{fig:spec_evol}), but on day 5.73, the contribution due to He I 6678 \AA\ becomes apparent. He I 6678 \AA\ is seen in the spectra till day 25.68.

He II 4686 \AA, 5412 \AA\ lines appear clearly on day 19.87 and persist throughout the epochs (see Figure \ref{fig:spec_evol}). They are also present in the accretion-dominated spectrum on day 147.66 (see Figure \ref{fig:7Nov_2021_spectra}). As discussed in Section \ref{EarlySpectra}, on day 5.73, He II 4686 \AA\ might be present on the right side of the feature centred at 4637 \AA, but the broad N III 4638 \AA\ line makes it difficult to confirm its presence (see Figure \ref{fig:He_II}). The He II 4686 \AA\ line's shape is different from the He I 7065 \AA, 5876 \AA\ and He II 5412 \AA\ lines. One reason for this could be the contribution due to other lines (N III 4638 \AA\ and [Ne IV] 4721 \AA), which are blended with the broad He II 4686 \AA\ line. Another reason could be the inherent non-uniformity of the ejecta, which can lead to such line profiles.

\begin{figure}
    \centering
    \includegraphics[width=\columnwidth]{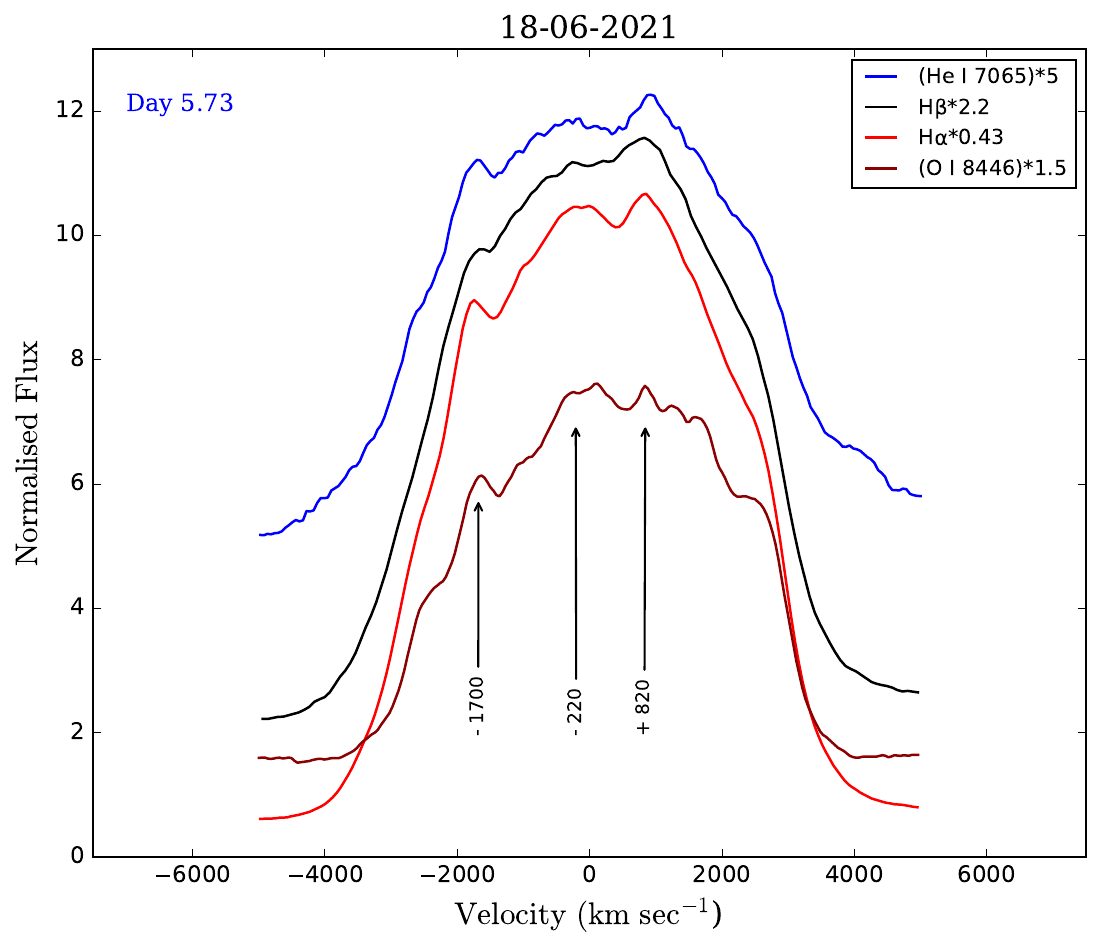}
    \caption{Velocity profiles of O I 8446 \AA, He I 7065 \AA, H$\alpha$, and H$\beta$ on day 5.73. The line fluxes are normalised to their respective local continuum values and then scaled using different constant values, as indicated in the legend.}
    \label{fig:OILineprofile_day5.73}
\end{figure}

\begin{figure}
    \centering
    \includegraphics[width=\columnwidth]{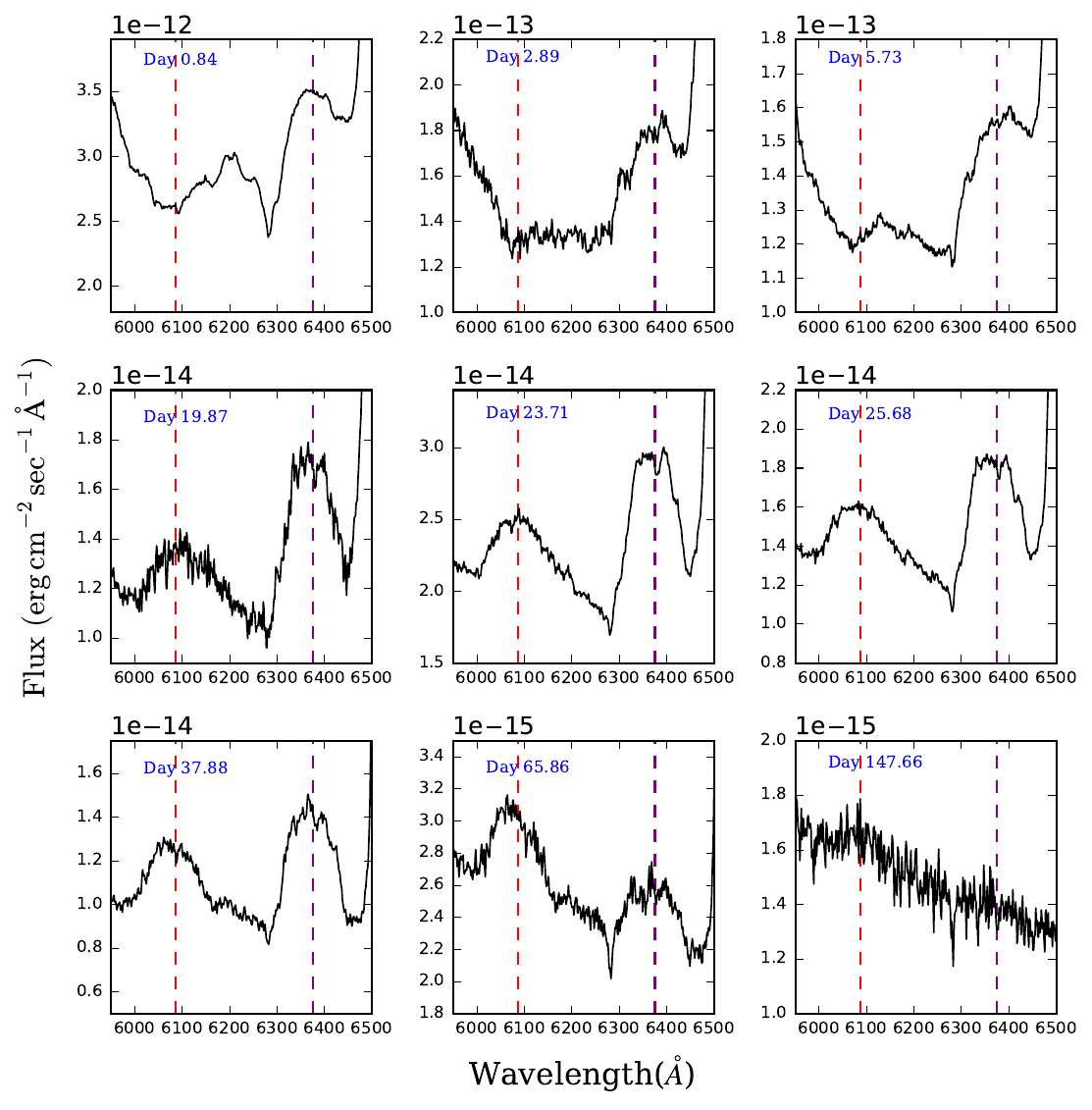}
    \caption{Evolution of [Fe VII] 6087 \AA\ \& [Fe X] 6375 \AA\ emission lines. Dotted lines represent the rest wavelengths of [Fe VII] 6087 \AA\ (red) \& [Fe X] 6375 \AA\ (purple). In the early spectra (day 0.84, 2.89, \& 5.73), [O I] 6364 \AA\ is possibly present. However, after the onset of the SSS phase on day 18.9, both [Fe VII] 6087 \AA\ and [Fe X] 6375 \AA\ lines were detected in the spectra on day 19.87 and remained present until day 65.86. On day 147.66, these lines are absent.}
    \label{fig:[FeVII_6087_FeX6375]}
\end{figure}

\begin{figure}
    \centering
    \includegraphics[width=\columnwidth]{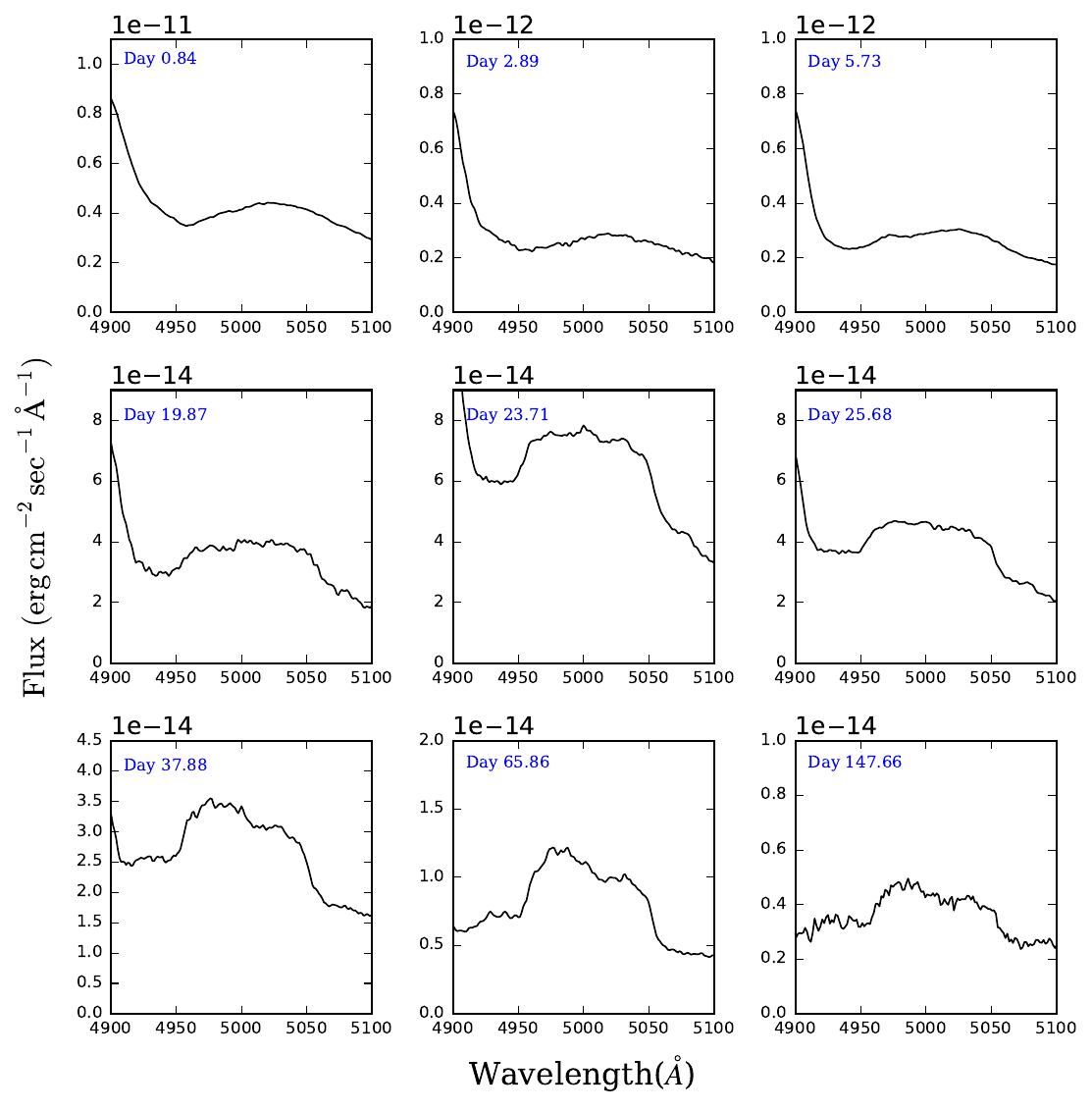}
    \caption{Evolution of the [O III] 4959 \AA\ \& 5007 \AA\ emission line profile across nine epochs. Initially, on day 0.84, N II 5001 \AA\ is present with possible contributions from He I 5016 \AA\ \& 5048 \AA\ as well as Fe II (42) multiplet at 5018 \AA. The [O III] 4959 \AA\ \& 5007 \AA\ emission line emerges on day 19.87 and persists until day 147.66.}
    \label{fig:[OIII]_variation}
\end{figure}

\subsection{Oxygen Lines} \label{Oxygen}
O I 8446 \AA, O I 7774 \AA, O I 8227 \AA\ in combination with N I 8212 \AA\ and O I 7477 \AA\ in combination with NI 7452 \AA\ are present in the optical spectrum on day 0.84. The strength of OI 8446 \AA\ is slightly higher than that of the O I 7774 \AA\ line. On day 2.89, there is an increase in the O I 8446 \AA\ line’s strength relative to OI 7774 \AA, with a ratio of O I 8446/7774 = 3.64 or O I 7774/8446 = 0.275. 

Neutral oxygen can be excited by recombination, continuum fluorescence, and Lyman beta fluorescence (\citet{1977ApJ...216...23S} and references therein). If recombination was to be the dominant mechanism, the ratio of O I 7774 \AA\ to O I 8446 \AA\ should have been 0.6. If continuum fluorescence was significant, the observed O I 1.3164 $\mu$m/1.1287 $\mu$m ratio would exceed unity. However, in \citet{2021ApJ...922L..10W}, on day 5.64, in the NIR spectra of V1674~Her, it is observed that the strength of the O I 1.3164 $\mu$m line is lower compared to O I 1.1287 $\mu$m. This indicates that Lyman beta fluorescence is the dominant excitation mechanism for O I 8446 \AA\ line and also means that the ejecta contains regions where both oxygen and hydrogen are neutral, and yet Lyman Beta flux density is high. This strengthening of the O I 8446 \AA\ continues till day 5.73, as shown in Figure \ref{fig:OI8446}, with a ratio of O I 8446/7774 = 7.93. The line profile of O I 8446 is also corrugated. In Figure \ref{fig:OILineprofile_day5.73}, the emission line profile of O I 8446 \AA\ for day 5.73 is plotted alongside H$\alpha$, H$\beta$, and He I 7065 \AA. The details in each profile exhibit remarkable similarities, particularly in the presence of identifiable features at radial velocities of -1700, -220, and 820 km/s. Subsequently, O I 8446 \AA\ starts to weaken, becoming weak on day 37.88 and absent on day 65.86.

Furthermore, the [O I] 6364 \AA\ line might be present in the initial epochs (see Figure \ref{fig:spec_evol}). However, in the later epochs, as the ejecta expands and the ionisation condition increases, the coronal and high ionisation lines of iron become dominant, resulting in the appearance of [Fe X] 6375 \AA\ line (see Figure \ref{fig:[FeVII_6087_FeX6375]}).

[O III] 4959 \AA, 5007 \AA\ lines first appear in the spectra on day 19.87 and are present throughout until day 147.66. The evolution of [O III] 4959 \AA, 5007 \AA\ lines is depicted in Figure \ref{fig:[OIII]_variation}. It is difficult to confirm the presence of the [O III] 4363 \AA\ line due to blending with H$\gamma$. However, the increase in the H$\gamma$/H$\beta$ flux on day 23.71 (see Table \ref{tab:flux_ratio}) indicates a secondary contribution due to some other line, which could possibly be [O III] 4363 \AA. The increase in flux is observed till day 65.86, making a case for the presence of the [O III] 4363 \AA\ line. On day 147.66 (see Figure \ref{fig:7Nov_2021_spectra}), [O III] 4363 \AA\ might be present, but it is difficult to confirm.

\begin{figure}
    \centering
    \includegraphics[width=\columnwidth]{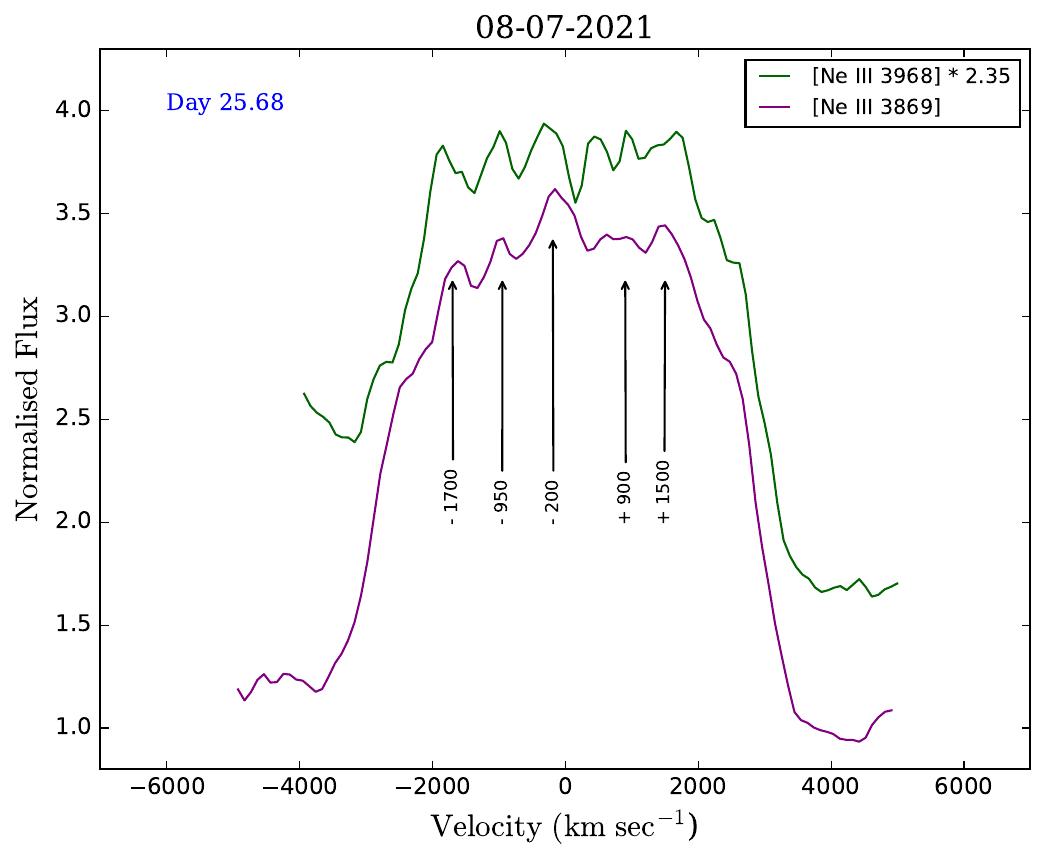}
    \caption{Velocity profiles of [Ne III] 3869 \AA, 3968 \AA\ on day 25.68. The line fluxes are normalised to their respective local continuum values. [Ne III] 3968 \AA\ line flux has been scaled by a constant number, as indicated in the legend.}
    \label{fig:Day_25.68_NeIII}
\end{figure}

\begin{figure*}
    \centering
    \includegraphics[width=2\columnwidth]{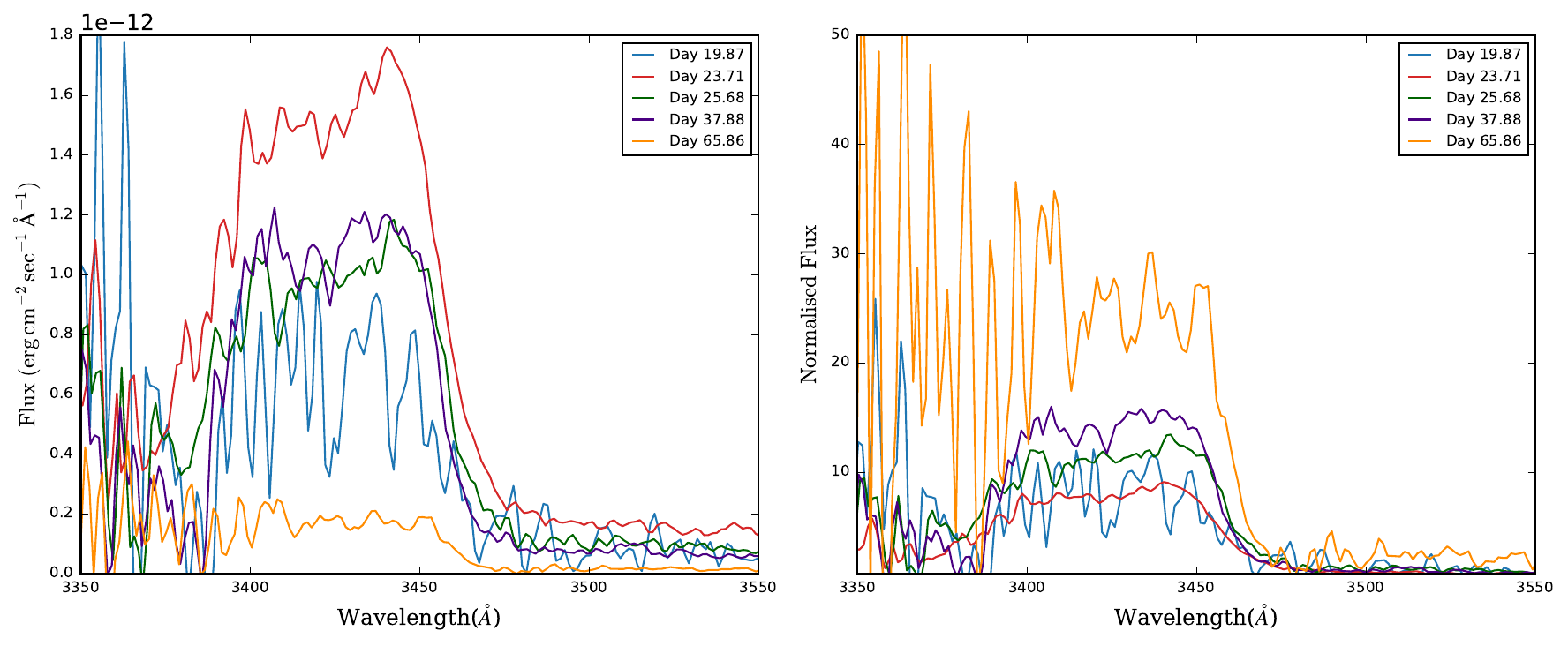}
    \caption{Evolution of the [Ne V] 3426 \AA\ line. The left panel shows the flux-calibrated spectra, while the right panel shows the same profiles, for each epoch, normalised to the local continuum. The [Ne V] 3426 \AA\ line is first detected on day 19.87, coinciding with the appearance of the [Ne III] 3869 \AA\ and 3968 \AA\ lines, and persists until day 65.86. On day 147.66, it is difficult to confirm the presence of [Ne V] 3426 \AA\ line (see Section \ref{Later_spectra}).}
    \label{fig:[NeV]_Variation}
\end{figure*}

\subsection{Neon Lines} \label{Neon}
In the ‘Early Spectra’ (see Section \ref{EarlySpectra}), hydrogen lines H$\zeta$ and H$\epsilon$ are observed. However, as the eruption progressed and transitioned into the SSS phase on day 18.9, as reported by \citet{2021ATel14747....1P}, neon lines, including [Ne III] 3869 \AA, [Ne III] 3968 \AA, and [Ne V] 3426 \AA, are observed in the spectra on day 19.87, indicating the presence of an ONe-type WD. 

Additionally, [Ne IV] 4721 \AA\ might also be present but is blended with the He II 4686 \AA\ line. The emission line profiles of both [Ne III] 3869 \AA\ and 3968 \AA\ are corrugated. Figure \ref{fig:Day_25.68_NeIII} shows the emission line profile of [Ne III] 3869 \AA\ and 3968 \AA\ for day 25.68, exhibiting similar features at radial velocities of -1700, -950, -200, 900, and 1500 km/s.

[Ne III] and [Ne V] lines persist in the spectra throughout until day 65.66. There is no significant change in the FWHM velocity of [Ne III] lines. The evolution of [Ne III] 3869 \AA\ and 3968 \AA\ lines is illustrated in Figure \ref{fig:[NeIII]_Variation}, and the evolution of [Ne V] 3426 \AA\ is depicted in Figure \ref{fig:[NeV]_Variation}. On day 147.66, Ne III 3869 \AA\ and 3968 \AA\ stand out after smoothing the spectra, although it is difficult to ascertain the presence of [Ne V] 3426 \AA\ line.

\begin{figure}
    \centering
    \includegraphics[width=\columnwidth]{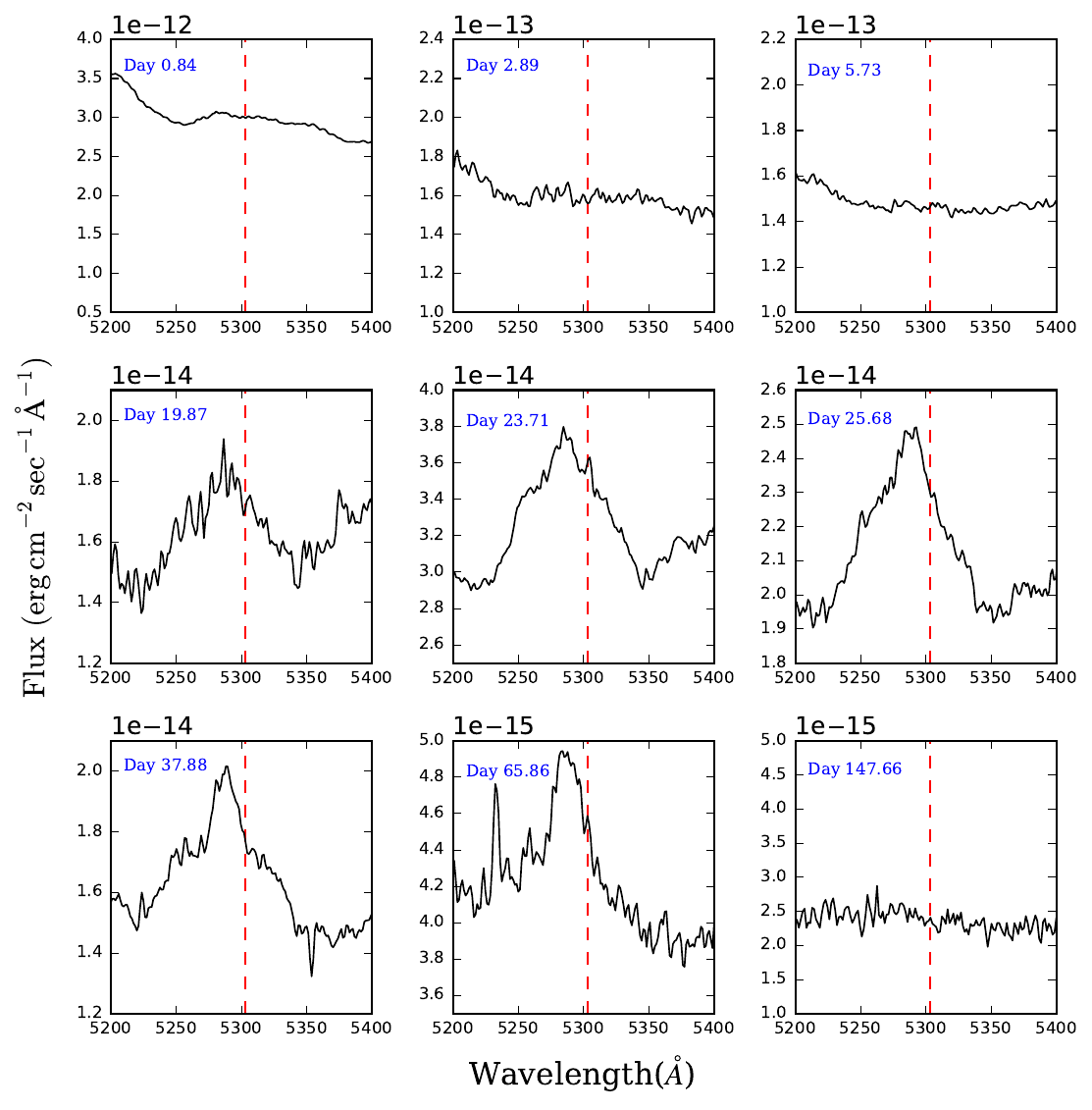}
    \caption{Evolution of [Fe XIV] 5303 \AA. Dotted line represents the rest wavelength of [Fe XIV] 5303 \AA. The emission feature became evident on day 19.87 and remained present until day 65.86. However, [Fe XIV] is absent on day 147.66.}
    \label{fig:[Fe XIV] 5303}
\end{figure} 

\begin{figure}
    \centering
    \includegraphics[width=\columnwidth]{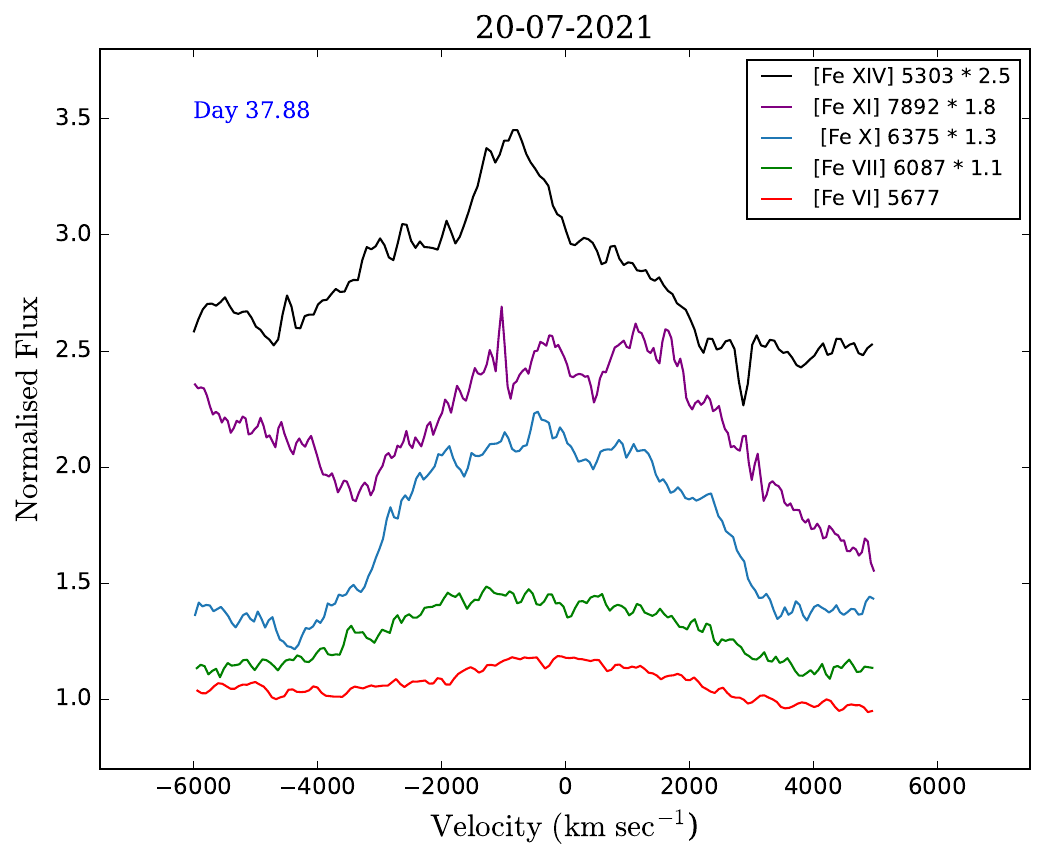}
    \caption{Velocity profiles of coronal and high ionisation lines of iron on day 37.88. The line fluxes are normalised to their respective local continuum values and then scaled by a constant number, as indicated in the legend.}
    \label{fig:coronal_lines}
\end{figure}

\begin{figure*}
    \centering
    \includegraphics[width=2\columnwidth]{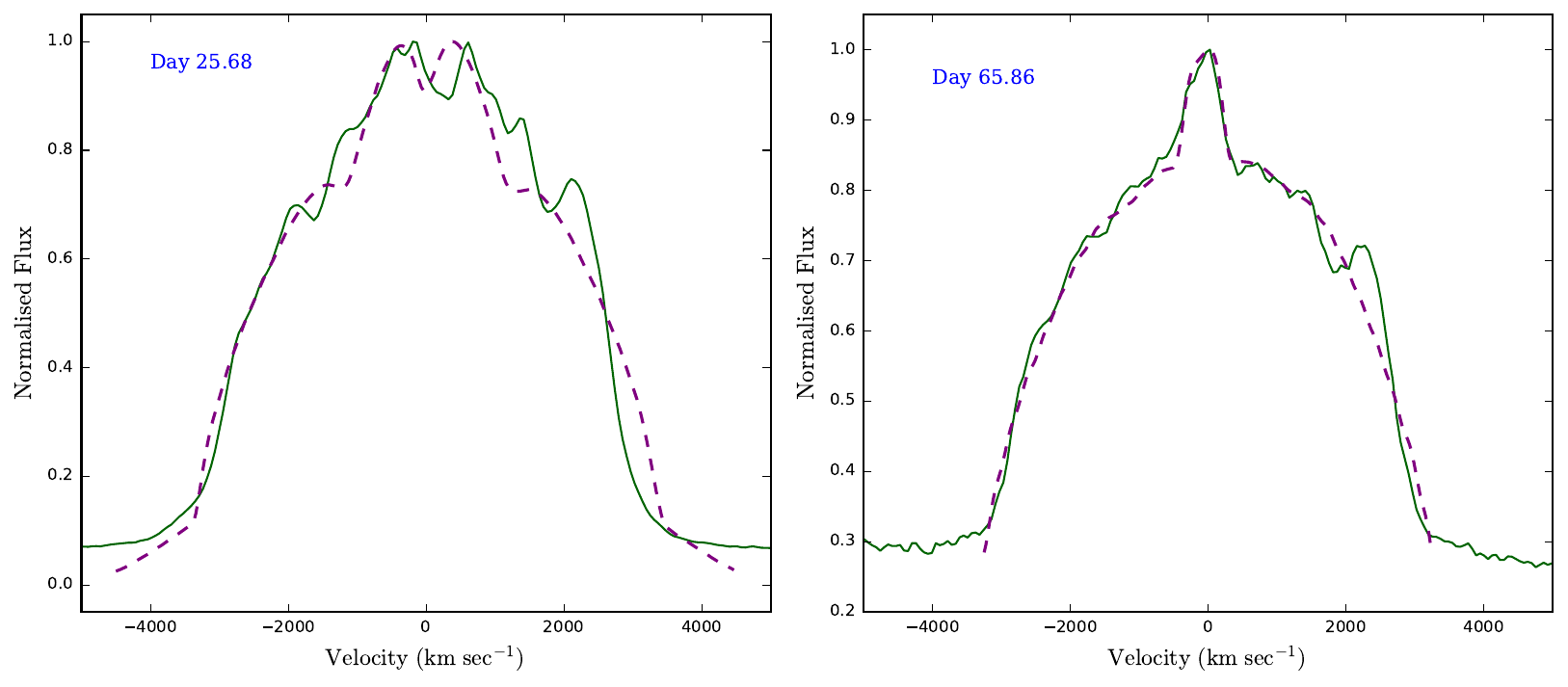}
    \caption{Observed (green solid line) and modelled (purple dashed line) H$\alpha$ line profiles of V1674~Her for day 25.68 (left) and day 65.86 (right). The velocity profiles are normalised to their respective peak flux values.}
    \label{fig:Model vs Observed_1}
\end{figure*} 

\subsection{High Ionisation and Coronal Lines}
On day 11.5, coronal lines (e.g. [Si VII] 1.96 $\mu$m, [Si VII] 2.48 $\mu$m) were observed in the NIR spectra of V1674~Her. \citet{2021ApJ...922L..10W} argued that the presence of these coronal lines could not have been caused by photoionisation, as the SSS phase was detected on day 18.9 \citep{2021ATel14747....1P}. Hence, shock ionisation was responsible for the presence of the coronal lines. We do not have data for the epoch between the shock initiation and the commencement of the SSS phase. Therefore, we cannot definitively comment on the shock contribution. In our case, we have spectra from day 19.87, which is after the commencement of the SSS phase, and hence, photoionisation is the dominant mechanism, though some elements of shock might be present. In the optical spectra of V1674~Her, [Fe VI] 5677 \AA, [Fe VII] 6087 \AA, [Fe X] 6375 \AA, [Fe XI] 7892 \AA, and [Fe XIV] 5303 \AA\ lines are observed.

On day 19.87, [Fe VI] 5677 \AA, [Fe VII] 6087 \AA, [Fe X] 6375 \AA, and [Fe XIV] 5303 \AA\ lines are present. Figure \ref{fig:[FeVI_5677]} illustrates the evolution of [Fe VI] 5677 \AA\ line. The line persists throughout the spectra till day 37.88. On day 65.86, it subsequently weakened and eventually disappeared on day 147.66. In Figure \ref{fig:[FeVII_6087_FeX6375]}, we observe the evolution of [Fe VII] 6087 \AA\ and [Fe X] 6375 \AA\ lines. The ‘Early spectra’ (see Section \ref{EarlySpectra}) potentially include the [O I] 6364 \AA\ line, but as the ionisation conditions increase, [Fe X] 6375 \AA\ is observed. Both [Fe VII] 6087 \AA\ and [Fe X] 6375 \AA\ persist until day 65.86 and are absent on day 147.66. [Fe XIV] 5303 \AA, appearing first on day 19.87, is present throughout until day 65.86 (see Figure \ref{fig:[Fe XIV] 5303}). On day 147.66, it is absent.

A weak emission line of [Fe XI] 7892 \AA\ is observed on day 23.71, becoming distinct on day 37.88. However, it is absent on day 65.86 (see Figure \ref{fig:[Fe XI] 7892}). Figure \ref{fig:coronal_lines} shows the emission line profiles of coronal and high ionisation lines of iron in the optical spectrum of V1674~Her on day 37.88.

\section{Analysis of H$\alpha$ Line Profile using \it SHAPE }\label{Modelling}
To investigate the geometry and understand the origin of the complex line profiles observed in V1674~Her, we performed a morpho-kinematic analysis of the ejecta using SHAPE. It is a morpho-kinematic tool that can be used to construct 3D models and generate a synthetic spectrum, which can then be compared to the observed spectrum. Although spatially resolved imaging can offer strong constraints when used in conjunction with spectral profiles, such data were not available at the time this modelling was initiated. Consequently, the modelling relied entirely on the spectral features.

\begin{figure*}
    \centering
    \includegraphics[width=2\columnwidth]{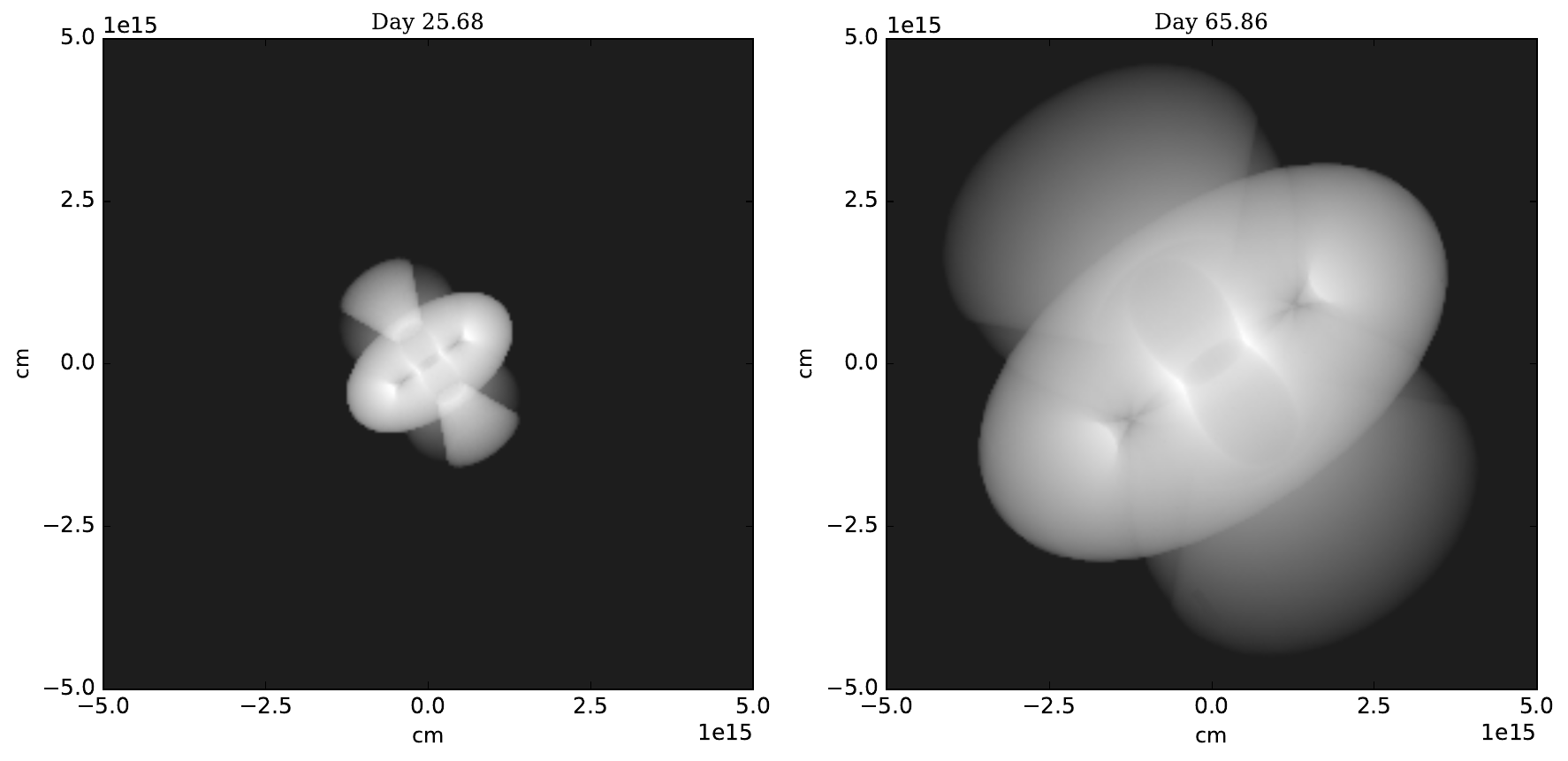}
    \caption{Morphology of the ejecta of V1674~Her using H$\alpha$ line profile for day 25.68 (left) and day 65.86 (right) in the 2d plane with x-axis being the line-of-sight direction and the y-axis being the axis perpendicular to that of the plane of sky and line-of-sight.}
    \label{fig:3dmorphology}
\end{figure*} 
The modelling was performed for two key epochs to capture the temporal evolution of the ejecta structure. For each epoch, the ejecta structure corresponding to the H$\alpha$ line profile was considered. SHAPE has been developed primarily for modelling optically thin environments. As discussed in section \ref{Oxygen}, in the June spectra, the excitation of neutral oxygen is found to be mainly driven by Lyman beta fluorescence, which indicates that the optical depth in H$\alpha$ was large at that stage. Therefore, Day 25.68 (2021 July 8) was selected as the first modelling epoch. This choice was also based on visual inspection of the H$\alpha$ profile, which then exhibited a more defined and structured morphology, making it suitable for geometric modelling. This epoch lies close to the latest epoch modelled by \citet{2024MNRAS.527.1405H}. To investigate the subsequent evolution of the ejecta, a second epoch was selected at day 65.86 (2021 August 17), providing a meaningful temporal baseline to assess morphological changes. This approach enables an assessment of whether the ejecta morphology inferred at earlier times remains consistent as the nova evolves. The other spectra obtained in July were deliberately left out so that the epochs would not be too closely spaced, allowing us to focus on broader structural development in the line profiles. For each epoch, we built 3D morpho-kinematic models, bringing together several geometric components to match the observed line shapes as closely as possible. The 3D module in SHAPE software provides a range of geometric primitive types, including sphere, torus, cone, cylinder, and plane, which can be combined and modified to construct complex ejecta structures. Using these tools, the initial geometry of the ejecta structure was created, starting with the assumption of a spheroidal shell featuring an equatorial ring and polar caps. The choice of these initial components was guided by earlier studies of nova shells (e.g. \cite{1999MNRAS.307..677G} and \citet{1972MNRAS.158..177H}). In addition to this starting configuration, we also explored simpler geometries constructed from individual primitives available within SHAPE to assess their ability to reproduce the observed H$\alpha$ line profiles. The radius of each component (or primitive) was calculated using the FWHM of the observed H$\alpha$ profile and the time since the eruption. Each primitive was assigned the necessary modifiers, such as density, velocity, and temperature. The density profile was assumed to vary with radius as $r^{-3}$, and the initial density value was taken from \citet{2024MNRAS.527.1405H}. A Hubble-flow type expansion was adopted for the velocity field \citep{2011PhDT.........1R}, expressed as
\begin{equation}
V_{\rm exp} = \frac{3200}{\sin i} \times \frac{r}{r_{0}},
\label{eq:vexp}
\end{equation}
where $r$ and $r_{0}$ are outer and inner radii, respectively, and $i$ is the inclination angle. These parameters were defined for the initial structure of ejecta. Additionally, the structural modifier squeeze was applied to shape the spherical shell into a bipolar form. The squeeze parameter represents the degree of shaping and describes the extent to which the structure is pinched along one axis. 
In our modelling, the parameter space for the angle of inclination was initially sampled over the full range of 0$\degree$– 90$\degree$. However, the primary modelling effort was concentrated within 55$\degree$– 75$\degree$, as discussed in Section \ref{Hydrogen}, where the inclination angle for V1674~Her is estimated to be around 60$\degree$ based on comparisons with the modelled profiles of \citet{1999MNRAS.307..677G} and the results of \citet{2024MNRAS.527.1405H}. Similarly, position angle was sampled over 0$\degree$– 180$\degree$. We found a position angle of 35$\degree$, consistent with the value reported by \citet{2024MNRAS.527.1405H}, to be most appropriate.

Once the initial geometry was set and its modifiers assigned, SHAPE was used to generate a synthetic 1D line profile for both epochs. During successive iterations, the values of these modifiers, along with the inclination angle, were adjusted to improve the match between the synthetic and observed profiles. In this modelling approach, the inclination angle is the primary parameter constrained by the H$\alpha$ line profiles. Other parameters, such as the density profile and velocity field, were adjusted qualitatively to reproduce the observed profile morphology. In the initial stages, the comparison was carried out visually to check whether the overall width and structure of the observed H$\alpha$ profiles were reasonably reproduced. Once the profiles seemed to match reasonably well, we calculated the quality of fit using the root mean square (rms) value and the quality factor (K). The rms value is given by
\begin{equation}
\mathrm{rms} = \sqrt{\frac{1}{N} \sum_{i=1}^{N} \left[ 1 - \frac{F_{\mathrm{model},i}}{F_{\mathrm{obs},i}} \right]^2 },
\label{eq:rms}
\end{equation}
where $N$ is the total number of velocity points, $F_{\mathrm{model},i}$ is the flux from the synthetic spectrum at the $i^{th}$ velocity point, and $F_{\mathrm{obs},i}$ is the corresponding flux from the observed spectrum. 

The quality factor was calculated following \citet{2009A&A...507.1517M}:
\begin{equation}
K = \frac{\log\left(\dfrac{F_{\mathrm{model}}}{F_{\mathrm{obs}}}\right)}{\log\left(1 + \sigma\right)}
\label{eq:K}
\end{equation}
where $F_{\mathrm{model}}$ and $F_{\mathrm{obs}}$ are the integrated model and observed fluxes, respectively, and $\sigma$ is the error in the observed line. A value of $|K| < 1$ indicates that the fit is within the chosen tolerance. For our analysis, we took $\sigma$ as 0.1, implying a conservative $10\%$ error in the observed flux.

The modelled H$\alpha$ line profiles for day 25.68 and day 65.86 are plotted in Figure \ref{fig:Model vs Observed_1}. The corresponding rms values are $0.2752$ and 0$.0498$, while the $K$ values are $-0.169$ and $-0.071$, respectively. The angle of inclination (i) is 65$\degree$. The morphology of the ejecta of V1674~Her using the H$\alpha$ line profile is shown in Figure \ref{fig:3dmorphology}. At both epochs, the observed H$\alpha$ line profiles are well reproduced by a morphology consisting of a bipolar shell, with polar blobs and an equatorial ring, similar to the configuration reported by \citet{2024MNRAS.527.1405H}. Simpler ejecta geometries, such as spherical shells or individual bipolar components, were unable to reproduce the observed corrugated structure and the multiple sub-peaks present in the H$\alpha$ profiles at these epochs.

This result ties into a broader discussion about how nova geometry relates to speed class. According to \citet{1995MNRAS.276..353S}, there is a relation between the speed class ($t_3$) of the nova and the axial ratio (the ratio of the lengths of the major and minor axes of the nova shell). They argued that faster novae should exhibit less pronounced shaping. However, as noted by \citet{2011MNRAS.412.1701R} and references therein, there are exceptions to this trend, such as V2672 Oph and RS Oph. In this context, the H$\alpha$ line profile modelling of V1674~Her is consistent with an ejecta morphology comprising a bipolar shell, polar blobs, and an equatorial ring, similar to those reported for some other fast novae.

The model further suggests that the central ring has a higher density in comparison to the surrounding squeezed shell. In Figure \ref{fig:3dmorphology}, we can notice that by the second epoch, i.e. day 65.86, the polar cones had broadened. This could possibly be a result of polar outflow not being collimated, allowing some material to spread sideways. Although the H$\alpha$ profiles are reproduced reasonably well, one must exercise necessary caution in over-interpreting both the results and geometry of the ejecta. 

Recently, \citet{article} presented early-time resolved imaging of the ejecta of V1674~Her obtained with CHARA within the first few days after eruption. Their observations revealed clear deviations from spherical symmetry, including evidence for multiple outflow components. Our SHAPE modelling for later epochs is broadly consistent with the overall morphology seen at those earlier epochs. More detailed modelling, incorporating other prominent line profiles or resolved imaging across multiple epochs, will be needed to place tighter constraints on the ejecta morphology. 

\section{Conclusions}\label{Results}
The evolution of the optical spectra and the light curves of the fast nova V1674~Herculis has been presented. We summarise our conclusions below:
\begin{enumerate}
  \item Using the AAVSO data, we find an orbital period of 0.153 days for V1674~Her, consistent with the previously reported value. No spin period is detected using the GIT data. The current GIT light curves are sensitive enough to detect spin modulations with an amplitude greater than 0.04 for the r$^\prime$ filter and 0.065 for the g$^\prime$ filter, indicating that the inherent optical pulsations are weaker than this.
   \item \(\textnormal{[Ne III]}\) and \(\textnormal{[Ne V]}\) lines are observed in the optical spectra, indicating the presence of an ONe white dwarf.
  \item In our optical spectra, coronal lines are initially observed on day 19.87 following the onset of the SSS phase, suggesting that photoionisation primarily drives the coronal line emission. However, it is possible that some contribution from shock ionisation exists. 
  \item The FWHM velocities of the H$\alpha$ and H$\beta$ lines are around 4500 to 5500 km/s for the first eight epochs. On day 147.66, the velocities are significantly different ($\sim1500$ km/s).
  \item On day 147.66, the coronal lines are absent, but [O III] nebular lines are still present, indicating that the nova has not yet returned to quiescence. He II 4686 \AA\ is observed with a much narrower FWHM ($\sim1500$ km/s) compared to the earlier epochs (>6000 km/s). This suggests that the line emission now originates from a different region, likely an accretion disk irradiated by the hot WD. The presence of the He II 4686 \AA\ line, along with the rising blue continuum, indicates an accretion-dominated spectrum.
  \item The strong O I 8446 \AA\ emission relative to O I 7774 \AA\ indicates that Lyman beta fluorescence is the dominant excitation mechanism for neutral oxygen.
  \item Using the H$\alpha$ line profiles, we see that the morphology of the V1674~Her ejecta consists of a bipolar shell with polar blobs and an equatorial ring, in agreement with the previously reported structures.
\end{enumerate}

\section*{Acknowledgements}
We thank the anonymous referee for their careful reading of the manuscript and for the constructive comments and suggestions which helped improve the clarity of this paper. We thank the staff of IAO, Hanle and Centre For Research \& Education in Science \& Technology (CREST), Hosakote, for help with the observations. The IAO and CREST facilities are operated by the Indian Institute of Astrophysics, Bangalore. We thank all the observers of HCT for accommodating some time for Target of Opportunity (ToO) observations. We also thank the HCT time allocation committee for the time and support during ToO and regular observations. The GROWTH India Telescope (GIT) is a 70-cm telescope with a 0.7-degree field of view, set up by the Indian Institute of Astrophysics (IIA) and the Indian Institute of Technology Bombay (IITB) with funding from  Indo-US Science and Technology Forum and the Science and Engineering Research Board, Department of Science and Technology, Government of India. It is located at the Indian Astronomical Observatory (Hanle), operated by IIA. We acknowledge funding by the IITB alumni batch of 1994, which partially supports operations of the telescope. The technical details of the telescope are available at \url{https://sites.google.com/view/growthindia/}. GCA and KPS thank the Indian National Science Academy for support under the Senior Scientist Programme. We gratefully acknowledge the contributions of observers worldwide to the AAVSO International Database, whose variable star observations were utilised in this research. This research used ASTROPY\footnote{\url{https://www.astropy.org}}, a community-developed core PYTHON package for Astronomy \citep{2018AJ....156..123A}. NSR sincerely thanks G.R. Habtie for kind help and constructive discussions on the SHAPE software.

%%%%%%%%%%%%%%%%%%%%%%%%%%%%%%%%%%%%%%%%%%%%%%%%%%
\section*{FACILITIES}
\textit{HCT, GIT, AAVSO}
%%%%%%%%%%%%%%%%%%%%%%%%%%%%%%%%%%%%%%%%%%%%%%%%%%
\section*{Data Availability}

HCT and GIT data will be provided by the corresponding author upon reasonable request. AAVSO data used for the analysis is available at \url{https://www.aavso.org/data-download}.


% Don't change these lines
\bsp	% typesetting comment
\label{lastpage}

\begin{thebibliography}{}
\makeatletter
\relax
\def\mn@urlcharsother{\let\do\@makeother \do\$\do\&\do\#\do\^\do\_\do\%\do\~}
\def\mn@doi{\begingroup\mn@urlcharsother \@ifnextchar [ {\mn@doi@} {\mn@doi@[]}}
\def\mn@doi@[#1]#2{\def\@tempa{#1}\ifx\@tempa\@empty \href {http://dx.doi.org/#2} {doi:#2}\else \href {http://dx.doi.org/#2} {#1}\fi \endgroup}
\def\mn@eprint#1#2{\mn@eprint@#1:#2::\@nil}
\def\mn@eprint@arXiv#1{\href {http://arxiv.org/abs/#1} {{\tt arXiv:#1}}}
\def\mn@eprint@dblp#1{\href {http://dblp.uni-trier.de/rec/bibtex/#1.xml} {dblp:#1}}
\def\mn@eprint@#1:#2:#3:#4\@nil{\def\@tempa {#1}\def\@tempb {#2}\def\@tempc {#3}\ifx \@tempc \@empty \let \@tempc \@tempb \let \@tempb \@tempa \fi \ifx \@tempb \@empty \def\@tempb {arXiv}\fi \@ifundefined {mn@eprint@\@tempb}{\@tempb:\@tempc}{\expandafter \expandafter \csname mn@eprint@\@tempb\endcsname \expandafter{\@tempc}}}

\bibitem[\protect\citeauthoryear{{Anupama} \& {Kamath}}{{Anupama} \& {Kamath}}{2012}]{2012BASI...40..161A}
{Anupama} G.~C.,  {Kamath} U.~S.,  2012, Bulletin of the Astronomical Society of India, \href {https://ui.adsabs.harvard.edu/abs/2012BASI...40..161A} {40, 161}

\bibitem[\protect\citeauthoryear{{Astropy Collaboration} et~al.,}{{Astropy Collaboration} et~al.}{2018}]{2018AJ....156..123A}
{Astropy Collaboration} et~al., 2018, \mn@doi [\aj] {10.3847/1538-3881/aabc4f}, \href {https://ui.adsabs.harvard.edu/abs/2018AJ....156..123A} {156, 123}

\bibitem[\protect\citeauthoryear{Aydi et~al.,}{Aydi et~al.}{2025}]{article}
Aydi E.,  et~al., 2025, \mn@doi [Nature Astronomy] {10.1038/s41550-025-02725-1}, pp 1--10

\bibitem[\protect\citeauthoryear{{Bhargava} et~al.,}{{Bhargava} et~al.}{2024}]{2024MNRAS.528...28B}
{Bhargava} Y.,  et~al., 2024, \mn@doi [\mnras] {10.1093/mnras/stad3870}, \href {https://ui.adsabs.harvard.edu/abs/2024MNRAS.528...28B} {528, 28}

\bibitem[\protect\citeauthoryear{{Drake} et~al.,}{{Drake} et~al.}{2021}]{2021ApJ...922L..42D}
{Drake} J.~J.,  et~al., 2021, \mn@doi [\apjl] {10.3847/2041-8213/ac34fd}, \href {https://ui.adsabs.harvard.edu/abs/2021ApJ...922L..42D} {922, L42}

\bibitem[\protect\citeauthoryear{{Gill} \& {O'Brien}}{{Gill} \& {O'Brien}}{1999}]{1999MNRAS.307..677G}
{Gill} C.~D.,  {O'Brien} T.~J.,  1999, \mn@doi [\mnras] {10.1046/j.1365-8711.1999.02681.x}, \href {https://ui.adsabs.harvard.edu/abs/1999MNRAS.307..677G} {307, 677}

\bibitem[\protect\citeauthoryear{{Habtie}, {Das}, {Pandey}, {Ashok}  \& {Dubovsky}}{{Habtie} et~al.}{2024}]{2024MNRAS.527.1405H}
{Habtie} G.~R.,  {Das} R.,  {Pandey} R.,  {Ashok} N.~M.,   {Dubovsky} P.~A.,  2024, \mn@doi [\mnras] {10.1093/mnras/stad3295}, \href {https://ui.adsabs.harvard.edu/abs/2024MNRAS.527.1405H} {527, 1405}

\bibitem[\protect\citeauthoryear{{Hutchings}}{{Hutchings}}{1972}]{1972MNRAS.158..177H}
{Hutchings} J.~B.,  1972, \mn@doi [\mnras] {10.1093/mnras/158.2.177}, \href {https://ui.adsabs.harvard.edu/abs/1972MNRAS.158..177H} {158, 177}

\bibitem[\protect\citeauthoryear{Kafka}{Kafka}{2021}]{Kafka2021}
Kafka S.,  2021, Observations from the AAVSO International Database. \url{https://www.aavso.org}

\bibitem[\protect\citeauthoryear{{Kawash} et~al.,}{{Kawash} et~al.}{2021}]{2021ApJ...910..120K}
{Kawash} A.,  et~al., 2021, \mn@doi [\apj] {10.3847/1538-4357/abe53d}, \href {https://ui.adsabs.harvard.edu/abs/2021ApJ...910..120K} {910, 120}

\bibitem[\protect\citeauthoryear{{Kumar} et~al.,}{{Kumar} et~al.}{2022a}]{2022AJ....164...90K}
{Kumar} H.,  et~al., 2022a, \mn@doi [\aj] {10.3847/1538-3881/ac7bea}, \href {https://ui.adsabs.harvard.edu/abs/2022AJ....164...90K} {164, 90}

\bibitem[\protect\citeauthoryear{{Kumar} et~al.,}{{Kumar} et~al.}{2022b}]{2022MNRAS.516.4517K}
{Kumar} H.,  et~al., 2022b, \mn@doi [\mnras] {10.1093/mnras/stac2516}, \href {https://ui.adsabs.harvard.edu/abs/2022MNRAS.516.4517K} {516, 4517}

\bibitem[\protect\citeauthoryear{{Lin}, {Fan}, {Hu}, {Takata}  \& {Li}}{{Lin} et~al.}{2022}]{2022MNRAS.517L..97L}
{Lin} L. C.-C.,  {Fan} J.-L.,  {Hu} C.-P.,  {Takata} J.,   {Li} K.-L.,  2022, \mn@doi [\mnras] {10.1093/mnrasl/slac117}, \href {https://ui.adsabs.harvard.edu/abs/2022MNRAS.517L..97L} {517, L97}

\bibitem[\protect\citeauthoryear{{Lomb}}{{Lomb}}{1976}]{lomb1976Ap&SS..39..447L}
{Lomb} N.~R.,  1976, \mn@doi [\apss] {10.1007/BF00648343}, \href {https://ui.adsabs.harvard.edu/abs/1976Ap&SS..39..447L} {39, 447}

\bibitem[\protect\citeauthoryear{{Luna}, {Lima}  \& {Orio}}{{Luna} et~al.}{2024}]{2024BAAA...65...60L}
{Luna} G.~J.~M.,  {Lima} I.~J.,   {Orio} M.,  2024, \mn@doi [Boletin de la Asociacion Argentina de Astronomia La Plata Argentina] {10.48550/arXiv.2310.02220}, \href {https://ui.adsabs.harvard.edu/abs/2024BAAA...65...60L} {65, 60}

\bibitem[\protect\citeauthoryear{{Maccarone}, {Beardmore}, {Mukai}, {Page}, {Pichardo Marcano}  \& {Rivera Sandoval}}{{Maccarone} et~al.}{2021}]{2021ATel14776....1M}
{Maccarone} T.~J.,  {Beardmore} A.,  {Mukai} K.,  {Page} K.,  {Pichardo Marcano} M.,   {Rivera Sandoval} L.,  2021, The Astronomer's Telegram, \href {https://ui.adsabs.harvard.edu/abs/2021ATel14776....1M} {14776, 1}

\bibitem[\protect\citeauthoryear{{Morisset} \& {Georgiev}}{{Morisset} \& {Georgiev}}{2009}]{2009A&A...507.1517M}
{Morisset} C.,  {Georgiev} L.,  2009, \mn@doi [\aap] {10.1051/0004-6361/200912413}, \href {https://ui.adsabs.harvard.edu/abs/2009A&A...507.1517M} {507, 1517}

\bibitem[\protect\citeauthoryear{{Mroz}, {Burdge}, {Roestel}, {Prince}, {Kong}  \& {Li}}{{Mroz} et~al.}{2021}]{2021ATel14720....1M}
{Mroz} P.,  {Burdge} K.,  {Roestel} J.~v.,  {Prince} T.,  {Kong} A.~K.~H.,   {Li} K.~L.,  2021, The Astronomer's Telegram, \href {https://ui.adsabs.harvard.edu/abs/2021ATel14720....1M} {14720, 1}

\bibitem[\protect\citeauthoryear{{Munari}, {Siviero}, {Dallaporta}, {Cherini}, {Valisa}  \& {Tomasella}}{{Munari} et~al.}{2011}]{2011NewA...16..209M}
{Munari} U.,  {Siviero} A.,  {Dallaporta} S.,  {Cherini} G.,  {Valisa} P.,   {Tomasella} L.,  2011, \mn@doi [\na] {10.1016/j.newast.2010.08.010}, \href {https://ui.adsabs.harvard.edu/abs/2011NewA...16..209M} {16, 209}

\bibitem[\protect\citeauthoryear{{Munari}, {Valisa}  \& {Dallaporta}}{{Munari} et~al.}{2021}]{2021ATel14704....1M}
{Munari} U.,  {Valisa} P.,   {Dallaporta} S.,  2021, The Astronomer's Telegram, \href {https://ui.adsabs.harvard.edu/abs/2021ATel14704....1M} {14704, 1}

\bibitem[\protect\citeauthoryear{{Nakano} et~al.,}{{Nakano} et~al.}{2008}]{2008IAUC.8934....1N}
{Nakano} S.,  et~al., 2008, \iaucirc, \href {https://ui.adsabs.harvard.edu/abs/2008IAUC.8934....1N} {8934, 1}

\bibitem[\protect\citeauthoryear{{Page}, {Orio}, {Sokolovsky}  \& {Kuin}}{{Page} et~al.}{2021}]{2021ATel14747....1P}
{Page} K.~L.,  {Orio} M.,  {Sokolovsky} K.~V.,   {Kuin} N.~P.~M.,  2021, The Astronomer's Telegram, \href {https://ui.adsabs.harvard.edu/abs/2021ATel14747....1P} {14747, 1}

\bibitem[\protect\citeauthoryear{{Patterson} et~al.,}{{Patterson} et~al.}{2022}]{2022ApJ...940L..56P}
{Patterson} J.,  et~al., 2022, \mn@doi [\apjl] {10.3847/2041-8213/ac9ebe}, \href {https://ui.adsabs.harvard.edu/abs/2022ApJ...940L..56P} {940, L56}

\bibitem[\protect\citeauthoryear{{Pei}, {Luna}, {Orio}, {Behar}, {Giese}, {Mikolajewska}  \& {Ness}}{{Pei} et~al.}{2021}]{2021ATel14798....1P}
{Pei} S.,  {Luna} G. J.~M.,  {Orio} M.,  {Behar} E.,  {Giese} M.,  {Mikolajewska} J.,   {Ness} J.-U.,  2021, The Astronomer's Telegram, \href {https://ui.adsabs.harvard.edu/abs/2021ATel14798....1P} {14798, 1}

\bibitem[\protect\citeauthoryear{{Quimby}, {Shafter}  \& {Corbett}}{{Quimby} et~al.}{2021}]{2021RNAAS...5..160Q}
{Quimby} R.~M.,  {Shafter} A.~W.,   {Corbett} H.,  2021, \mn@doi [Research Notes of the American Astronomical Society] {10.3847/2515-5172/ac14c0}, \href {https://ui.adsabs.harvard.edu/abs/2021RNAAS...5..160Q} {5, 160}

\bibitem[\protect\citeauthoryear{{Ribeiro}}{{Ribeiro}}{2011}]{2011PhDT.........1R}
{Ribeiro} V. A.~R.~M.,  2011, PhD thesis, Liverpool John Moores University, UK

\bibitem[\protect\citeauthoryear{{Ribeiro}, {Darnley}, {Bode}, {Munari}, {Harman}, {Steele}  \& {Meaburn}}{{Ribeiro} et~al.}{2011}]{2011MNRAS.412.1701R}
{Ribeiro} V.~A.~R.~M.,  {Darnley} M.~J.,  {Bode} M.~F.,  {Munari} U.,  {Harman} D.~J.,  {Steele} I.~A.,   {Meaburn} J.,  2011, \mn@doi [\mnras] {10.1111/j.1365-2966.2010.18006.x}, \href {https://ui.adsabs.harvard.edu/abs/2011MNRAS.412.1701R} {412, 1701}

\bibitem[\protect\citeauthoryear{{Scargle}}{{Scargle}}{1982}]{scargle1982ApJ...263..835S}
{Scargle} J.~D.,  1982, \mn@doi [\apj] {10.1086/160554}, \href {https://ui.adsabs.harvard.edu/abs/1982ApJ...263..835S} {263, 835}

\bibitem[\protect\citeauthoryear{{Shugarov} \& {Afonina}}{{Shugarov} \& {Afonina}}{2021}]{2021ATel14835....1S}
{Shugarov} S.,  {Afonina} M.,  2021, The Astronomer's Telegram, \href {https://ui.adsabs.harvard.edu/abs/2021ATel14835....1S} {14835, 1}

\bibitem[\protect\citeauthoryear{{Slavin}, {O'Brien}  \& {Dunlop}}{{Slavin} et~al.}{1995}]{1995MNRAS.276..353S}
{Slavin} A.~J.,  {O'Brien} T.~J.,   {Dunlop} J.~S.,  1995, \mn@doi [\mnras] {10.1093/mnras/276.2.353}, \href {https://ui.adsabs.harvard.edu/abs/1995MNRAS.276..353S} {276, 353}

\bibitem[\protect\citeauthoryear{{Starrfield}, {Iliadis}, {Timmes}, {Hix}, {Arnett}, {Meakin}  \& {Sparks}}{{Starrfield} et~al.}{2012}]{2012BASI...40..419S}
{Starrfield} S.,  {Iliadis} C.,  {Timmes} F.~X.,  {Hix} W.~R.,  {Arnett} W.~D.,  {Meakin} C.,   {Sparks} W.~M.,  2012, \mn@doi [Bulletin of the Astronomical Society of India] {10.48550/arXiv.1210.6086}, \href {https://ui.adsabs.harvard.edu/abs/2012BASI...40..419S} {40, 419}

\bibitem[\protect\citeauthoryear{{Starrfield}, {Bose}, {Iliadis}, {Hix}, {Woodward}  \& {Wagner}}{{Starrfield} et~al.}{2020}]{Starrfield2020ApJ...895...70S}
{Starrfield} S.,  {Bose} M.,  {Iliadis} C.,  {Hix} W.~R.,  {Woodward} C.~E.,   {Wagner} R.~M.,  2020, \mn@doi [\apj] {10.3847/1538-4357/ab8d23}, \href {https://ui.adsabs.harvard.edu/abs/2020ApJ...895...70S} {895, 70}

\bibitem[\protect\citeauthoryear{{Steffen}, {Koning}, {Wenger}, {Morisset}  \& {Magnor}}{{Steffen} et~al.}{2011}]{2011ITVCG..17..454S}
{Steffen} W.,  {Koning} N.,  {Wenger} S.,  {Morisset} C.,   {Magnor} M.,  2011, \mn@doi [IEEE Transactions on Visualization and Computer Graphics] {10.1109/TVCG.2010.62}, \href {https://ui.adsabs.harvard.edu/abs/2011ITVCG..17..454S} {17, 454}

\bibitem[\protect\citeauthoryear{{Strittmatter} et~al.,}{{Strittmatter} et~al.}{1977}]{1977ApJ...216...23S}
{Strittmatter} P.~A.,  et~al., 1977, \mn@doi [\apj] {10.1086/155438}, \href {https://ui.adsabs.harvard.edu/abs/1977ApJ...216...23S} {216, 23}

\bibitem[\protect\citeauthoryear{{VanderPlas}}{{VanderPlas}}{2018}]{vdp2018ApJS..236...16V}
{VanderPlas} J.~T.,  2018, \mn@doi [\apjs] {10.3847/1538-4365/aab766}, \href {https://ui.adsabs.harvard.edu/abs/2018ApJS..236...16V} {236, 16}

\bibitem[\protect\citeauthoryear{{Wagner}, {Woodward}, {Starrfield}, {Banerjee}  \& {Evans}}{{Wagner} et~al.}{2021}]{2021ATel14746....1W}
{Wagner} R.~M.,  {Woodward} C.~E.,  {Starrfield} S.,  {Banerjee} D.~P.~K.,   {Evans} A.,  2021, The Astronomer's Telegram, \href {https://ui.adsabs.harvard.edu/abs/2021ATel14746....1W} {14746, 1}

\bibitem[\protect\citeauthoryear{{Warner}}{{Warner}}{1995}]{1995cvs..book.....W}
{Warner} B.,  1995, {Cataclysmic variable stars}.
 Vol. 28

\bibitem[\protect\citeauthoryear{{Williams}}{{Williams}}{1992}]{1992AJ....104..725W}
{Williams} R.~E.,  1992, \mn@doi [\aj] {10.1086/116268}, \href {https://ui.adsabs.harvard.edu/abs/1992AJ....104..725W} {104, 725}

\bibitem[\protect\citeauthoryear{{Williams}}{{Williams}}{2012}]{2012AJ....144...98W}
{Williams} R.,  2012, \mn@doi [\aj] {10.1088/0004-6256/144/4/98}, \href {https://ui.adsabs.harvard.edu/abs/2012AJ....144...98W} {144, 98}

\bibitem[\protect\citeauthoryear{{Woodward}, {Banerjee}, {Geballe}, {Page}, {Starrfield}  \& {Wagner}}{{Woodward} et~al.}{2021a}]{2021ApJ...922L..10W}
{Woodward} C.~E.,  {Banerjee} D.~P.~K.,  {Geballe} T.~R.,  {Page} K.~L.,  {Starrfield} S.,   {Wagner} R.~M.,  2021a, \mn@doi [\apjl] {10.3847/2041-8213/ac3518}, \href {https://ui.adsabs.harvard.edu/abs/2021ApJ...922L..10W} {922, L10}

\bibitem[\protect\citeauthoryear{{Woodward} et~al.,}{{Woodward} et~al.}{2021b}]{2021ATel14723....1W}
{Woodward} C.~E.,  et~al., 2021b, The Astronomer's Telegram, \href {https://ui.adsabs.harvard.edu/abs/2021ATel14723....1W} {14723, 1}

\bibitem[\protect\citeauthoryear{{Woodward}, {Banerjee}, {Wagner}, {Starrfield}  \& {Evans}}{{Woodward} et~al.}{2021c}]{2021ATel14728....1W}
{Woodward} C.~E.,  {Banerjee} D.~P.~K.,  {Wagner} R.~M.,  {Starrfield} S.,   {Evans} A.,  2021c, The Astronomer's Telegram, \href {https://ui.adsabs.harvard.edu/abs/2021ATel14728....1W} {14728, 1}

\bibitem[\protect\citeauthoryear{{Woodward}, {Banerjee}, {Evans}, {Wagner}  \& {Starrfield}}{{Woodward} et~al.}{2021d}]{2021ATel14741....1W}
{Woodward} C.~E.,  {Banerjee} D.~P.~K.,  {Evans} A.,  {Wagner} R.~M.,   {Starrfield} S.,  2021d, The Astronomer's Telegram, \href {https://ui.adsabs.harvard.edu/abs/2021ATel14741....1W} {14741, 1}

\bibitem[\protect\citeauthoryear{{Yaron}, {Prialnik}, {Shara}  \& {Kovetz}}{{Yaron} et~al.}{2005}]{2005ApJ...623..398Y}
{Yaron} O.,  {Prialnik} D.,  {Shara} M.~M.,   {Kovetz} A.,  2005, \mn@doi [\apj] {10.1086/428435}, \href {https://ui.adsabs.harvard.edu/abs/2005ApJ...623..398Y} {623, 398}

\makeatother
\end{thebibliography}
\end{document}